\documentclass[12pt,a4paper,final]{iopart}
\usepackage{iopams}
\usepackage{graphicx}
\usepackage[breaklinks=true,colorlinks=true,linkcolor=blue,urlcolor=blue,citecolor=blue]{hyperref}

\begin{document}

\title{Soliton nanoantennas in two-dimensional arrays of quantum dots}
\author{G. Gligori\'{c}, A. Maluckov, and Lj. Had\v{z}ievski}
\address{P* group, Vin\v{c}a Institute of Nuclear Sciences, University of
Belgrade, P. O. B. 522,11001 Belgrade, Serbia}
\author{ G. Ya. Slepyan, and B. A. Malomed}
\address{Department of Physical Electronics, School of Electrical
Engineering, Faculty of Engineering, Tel Aviv University, Tel Aviv
69978, Israel}\ead{sandram@vin.bg.ac.rs}

\begin{abstract}
We consider two-dimensional (2D) arrays of self-organized semiconductor
quantum dots (QDs) strongly interacting with electromagnetic field in the
regime of Rabi oscillations. The QD array built of two-level states is
modelled by two coupled systems of discrete nonlinear Schr\"{o}dinger
equations. Localized modes in the form of single-peaked fundamental and
vortical stationary Rabi solitons and self-trapped breathers have been
found. The results for the stability, mobility and radiative properties of
the Rabi modes suggest a concept of a self-assembled 2D \textit{%
soliton-based nano-antenna}, which should be stable against imperfections In
particular, we discuss the implementation of such a nano-antenna in the form
of surface plasmon solitons in graphene, and illustrate possibilities to
control their operation by means of optical tools.
\end{abstract}


\address{P* group, Vin\v{c}a Institute of Nuclear Sciences, University of
Belgrade, P. O. B. 522,11001 Belgrade, Serbia}

\address{Department of Physical Electronics, School of Electrical
Engineering, Faculty of Engineering, Tel Aviv University, Tel Aviv
69978, Israel}\ead{sandram@vin.bg.ac.rs}


\vspace{2.0075pc} \noindent \textit{Keywords}: two-dimensional dynamical
lattices, Rabi solitons, discrete solitons, quantum dots, nano-antenna

\section{Introduction}

Experimental and theoretical studies of solitons in nonlinear lattices are
the field of intensive activity in several branches of physics. Its large
part is focused on nonlinear optics \cite{optics-review} and matter waves
(i.e., Bose-Einstein condensates, BEC). The latter may be treated in the
mean-field approximation \cite{BEC-original}, or in terms of the quantum
Bose-Hubbard model \cite{BH,BH-review} (see book \cite{Maciek} for a
systematic summary). In these contexts, the discrete nonlinear Schr\"{o}%
dinger (DNLS) equation, and systems of the coupled DNLS equations, are
ubiquitous dynamical models for the description of nonlinear lattices \cite%
{1}. In the framework of these studies, the existence, stability, and
dynamics (including mobility) of discrete self-trapped modes (discrete
solitons) is a topic which has drawn a great deal of interest.

In multi-component systems, couplings between components may be linear,
nonlinear, or both linear and nonlinear \cite{1}. In the context of optics,
systems of linearly coupled DNLS equations are relevant to various
applications. In particular, the linear coupling between two polarization
modes in each core of a waveguide array may be induced by a twist of the
core (for linear polarizations), or by the birefringence (for circular
polarizations). On the other hand, in the BEC the linear coupling may be
imposed by an external microwave or radio-frequency field, which can drive
Rabi oscillations \cite{5,6} or Josephson oscillations \cite{7,8} between
two boson populations. In terms of the dynamical analysis, an essential
issue is the possibility of the spontaneous symmetry breaking in linearly
coupled discrete systems \cite{Herring}.

The progress in the fabrication of nano-scale electric circuits
\cite{16}, lasers \cite{17}, waveguides and antennas, for
microwave, tera-Hertz and optical frequency ranges, is another
strong incentive stimulating studies of nonlinear lattices.
Particularly significant nano-dimensional elements are
self-organized quantum-dots (QDs) \cite{18} embedded into
semiconductor dielectric host media. The possibility of strong
interactions of QDs with optical fields gives rise to Rabi
oscillations (RO) between electron-hole populations of ground and
excited states. The basic RO\ model amounts to a two-level atom
strongly coupled to an external electromagnetic field (the Jaynes
- Cummings model) \cite{19}. For more complex structures, such as
large molecules \cite{20} or QDs, this model has been modified to
account for additional factors: anisotropy \cite{20}, local-field
corrections \cite{21}-\cite{24}, broken inversion symmetry
\cite{25}, etc. In these settings, crucially significant are
collective coherent effects, stipulated by the inter-dot coupling
inside the QD array. Oscillations of the QD population between
levels in an isolated two-level system may be considered as energy
exchange between the system and the ambient electromagnetic field.
On the other hand, particle transfer (e.g., inter-dot electron, or
electron-hole tunnelling) leads to the exchange of the
quasi-momentum between charge carriers and photons. These
mechanisms govern the spatiotemporal RO dynamics in the QD chains,
in the form of the propagation of travelling RO waves and wave
packets (\textit{Rabi waves}) \cite{26}-\cite{28}.

In the QDs arrays the transport can also be realized by tunnelling via the
inter-dot dipole-dipole coupling, such as F\"{o}rster interactions \cite{18}%
, and by the radiation-field transfer. In Ref. \cite{27}, the tunnelling was
assumed to be a dominant mechanism of the inter-dot coupling. In that case,
a significant role is played by local-field effects in the QDs, even in the
weak-coupling regime \cite{21,22}. The local-field effects, enhanced by the
strong light-QD coupling, modify the conventional RO dynamics \cite{26}-\cite%
{28}. As a result, specific nonlinear terms appear in equations of
electron-hole oscillations inside the QDs, braking the superposition
principle for the single-particle wave function. The role of local fields in
the formation of excitonic RO in self-assembling QDs was experimentally
studied in Refs. \cite{29,30}.

The nonlinearity affects the Rabi-wave propagation too. As was shown in Ref.
\cite{31}, solitons and self-trapped breathers in the one-dimensional (1D)
QD chains can be created with the help of the nonlinearity. The model
developed in Ref. \cite{31} explores the dynamics of the probability
amplitudes of the ground and excited states of QDs, with included nonlinear
cross-phase-modulation terms and linear coupling between the two states. The
phase shift between constants of the inter-site hopping is a new property
introduced geometrically in that 1D setting.

Recently, theoretical and experimental studies of nano-antennas have been in
a focus of work performed by many groups. The antenna is defined as a device
which transforms the near field into the far field, or vice versa
(transmitting and receiving antennas, respectively). Transmitting devices
produce the spherical wave in a remote spatial area, with electric field
\begin{equation}
\mathbf{E}_{\mathrm{Rad}}=\frac{\mathbf{e}_{\theta }}{4\pi \epsilon _{0}}%
\frac{\omega ^{2}}{c^{2}}\mathbf{E}_{0}\frac{e^{-i\omega (t-R/c)}}{R}%
F(\theta ,\varphi )+c.c.,  \label{eq1}
\end{equation}%
where $R,\theta ,\varphi $ are the spherical coordinates, with the origin
set at the center of the antenna, $F(\theta ,\varphi )$ is a normalized
angular radiation pattern, $\epsilon _{0}$ is the vacuum permittivity, $c$
is the free-space light velocity, $\mathbf{e}_{\theta }$ is the unit vector
along the local direction of coordinate $\theta $, with $\mathbf{E}_{0}$ and
$\omega $ being the amplitude and frequency of the electric field,
respectively. The purport of using the antennas is their ability to provide
an interface between local information processing, which uses electrical
signals, and the free-space wireless transmission of data encoded in various
parameters of the electromagnetic waves, such as the amplitude, phase and
frequency.

Directional antenna properties (the dependence of the emitted
field on both the azimuthal and polar angles) are characterized by
the radiation pattern. It is determined by the near-field
distribution produced by the signal source placed inside the
antenna. It depends on the frequency, antenna configuration and
geometric parameters. An important species of the device is
represented by phased antenna arrays, i.e., systems of a large
number of identical emitters with a phase shift between adjacent
ones \cite{32}. The progress in the fabrication of nano-antennas,
see Refs. \cite{33}-\cite{35} for review, manifests the general
trend of implementing radio-communication principles in the
optical-frequency range. In particular, it stimulates bridging the
realms of macroscopic antennas and nano-antennas, using new
materials with specific electronic properties. In this context, it
is relevant to mention carbon nanotubes and nanotube arrays
\cite{36}-\cite{39}, plasmonic noble-metal wires \cite{40,41},
graphene nanoribbons \cite{42}, semiconductor QDs \cite{43,44},
etc. These types of nano-antennas have been offered as promising
elements for industrial design and commercial manufacturing.
However, many limitations for their use are stipulated by
difficulties of their operational control -- for instance,
rotation of the radiation pattern by an electric or optical drive.
This difficulty, along with the growing interest to the field,
motivate the search of new physical principles for fabrication and
tuning of nano-antennas. A promising possibility for that is
suggested by using strong nonlinearities in nanostructures (in
particular, remarkable nonlinear properties of graphene
\cite{3new,4new}).

In this work, we propose a previously unexplored principle for the
realization of nano-antenna arrays in the form of self-trapping \emph{%
discrete solitons }in 2D nonlinear lattices built of semiconductor QDs. In
this context, dynamically stable soliton structures (fundamental discrete
solitons, vortex solitons, and breathers) may provide promising mechanisms
of controllable creation of the antennas with various geometric shapes. The
model for the 2D array of QDs can be developed from its 1D counterpart which
was elaborated in Ref. \cite{31}; obviously, the transition from 1D to the
2D geometry is a crucially important step for modelling antennas which emit
in the plane of the semiconductor structure. We demonstrate that different
types of the global drive, \textit{viz}., plane and cylindrical waves, can
excite Rabi oscillations and waves in the system. Using known numerical
techniques developed for the study of discrete solitons \cite{1}, we
construct basic localized modes of the corresponding 2D DNLS equations and
test their stability and mobility. In addition, we demonstrate how discrete
Rabi solitons induce the dielectric polarization, and thus build the
near-field of the soliton-based nano-antenna. On the contrary to other
radiation zones, the near field keeps itself in a quantum state, which helps
to establish the correlation of quantum effects and field transformation
between different spatial zones. We also present the far-field structure
produced by the given near field, which takes into account the quantum
nature of the underlying Rabi oscillator.

The paper is organized as follows. In Sec. II, we introduce the model for
the 2D QD array interacting with the driving electromagnetic field. As
mentioned above, the plane- and cylindrical-wave drives are considered. The
nonlinearity is induced as the local-field effect. The model is analyzed
numerically, with the purpose to reveal the RO dynamics in the QD arrays. In
Sec. III, necessary technical ingredients are introduced, including the
dispersion relation for the linearized system, and general expressions for
radiation properties generated by solitons in the 2D QD array. In Sec. IV,
results are presented in detail, and the relation of the present model to 2D
nano-antenna arrays is proposed. The important issue of maintaining the
coherence of emitters building the soliton nano-antenna is considered in
Sec. IV too, and possibilities for operational control of the device by
means of optical tools are outlined. The paper is summarized in Sec. V.

\section{Two-dimensional model equations}

\subsection{The plane-wave drive}

\begin{figure}[h]
\center\includegraphics[width=13cm]{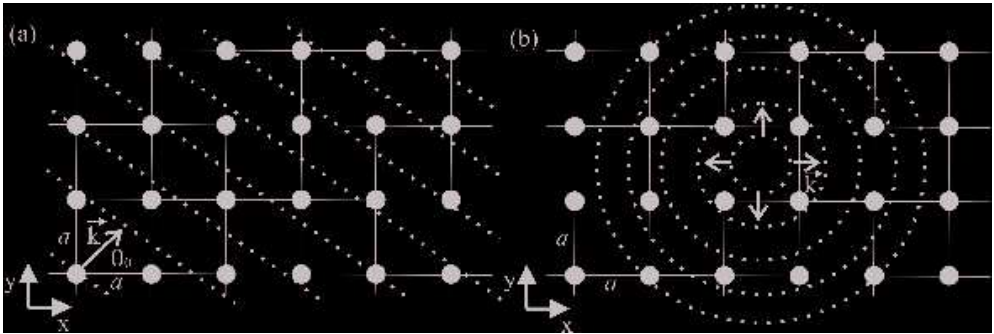} \caption{
Schematically presented plane wave (a) and cylindrical wave (b)
drive of 2D QD array placed in the $(x,y)$ plane. Dotted curves
display wave fronts in both cases, $\vec{k}$ is the wave-vector,
$\theta_0$ is the propagation angle and $a$ is the inter-dot
distance. } \label{waves}
\end{figure}

We consider finite 2D rectangular $N_{1}\times N_{2}$ array of identical QDs
with the square unit cell of size $a$ $\times $ $a$ ($a$ is the inter-dot
distance). The position of QDs is defined by a pair of discrete indices, $%
p=-N_{1}/2,...,N_{1}/2$ and $q=-N_{2}/2,...,N_{2}/2$. The array is exposed
to the classical travelling optical wave of in the $xy$- plane with electric
field%
\begin{equation}
\mathbf{E}(x,y,t)\mathbf{=}\mathrm{Re}\left\{ \mathbf{E}_{0}\exp \left[
ik\left( x\cos \theta _{0}+y\sin \theta _{0}\right) -i\omega t\right]
\right\} ,  \label{eqe}
\end{equation}
where $k=k(\omega )$ is the wavenumber and $\theta _{0}$ is the
propagation angle (see Fig. \ref{waves} (a)). The driving field
may represent different types of the electromagnetic modes, the
first one being a plane wave with a nonzero $E_{z} $ component.
Another example is a surface plasmon guided by the plane boundary
between the noble metal and transparent medium \cite{18}, on which
the QD array is installed. The latter example comes with the
frequency dispersion of a plasma medium, which manifests itself in
a nonlinear dependence $k=k(\omega )$. We take QDs as identical
two-level dissipationless systems with the transition energy,
$\hbar \omega _{0}$, corresponding to the transition between
excited and ground-state electron orbitals, $|a_{p,q}\rangle $ and
$\,|b_{p,q}\rangle $. The transition dipolar moment is
$\mathbf{\mu }=\mathbf{e}_{z}\mu $, and only intraband transitions
are taken into account \cite{27}. Each QD is coupled by the
electron tunnelling to four nearest neighbors in the 2D array.

It is adopted that the light interaction within the QD chain takes place in
the resonant regime, i.e., the frequency detuning is small in comparison to
both the light and quantum-transition frequencies. Following the
rotation-wave-approximation \cite{19}, we omit rapidly oscillating terms in
the equations of motion. These natural assumptions follows the Rabi-waves
model formulated in Ref. \cite{27}.

The raising, lowering, and population operators of the QD by denoted as $%
\hat{\sigma}_{p,q}^{+}=|a_{p,q}\rangle \langle b_{p,q}|,\,\hat{\sigma}%
_{p,q}^{-}=|b_{p,q}\rangle \langle a_{p,q}|,$ and $\hat{\sigma}%
_{zp,q}=|a_{p,q}\rangle \langle a_{p,q}|-|b_{p,q}\rangle \langle b_{p,q}|$,
respectively. The total Hamiltonian is
\begin{equation}
\hat{H}=\hat{H}_{d}+\hat{H}_{df}+\hat{H}_{T}+\Delta \hat{H},  \label{eq2}
\end{equation}%
where the term
\begin{equation}
\hat{H}_{d}=\frac{\hbar \omega _{0}}{2}\sum_{p,q}{\hat{\sigma}_{zp,q}}
\label{eq3}
\end{equation}%
describes the free electron motion, while the one
\begin{equation}
\hat{H}_{df}=-(\mathbf{\mu E}_{0})\sum_{p,q}\hat{\sigma}_{p,q}^{+}\exp {%
(i(p\phi _{1}+q\phi _{2}))}+\mathrm{H.c}.  \label{eq4}
\end{equation}%
($\mathrm{H.c.}$ stands for Hermitian-conjugate operator) accounts for the
interaction of QD array with the electromagnetic field in the dipole
approximation without tunnelling. Here, phases
\begin{equation}
\phi _{1}=(ka/2)\cos {(\theta )},\phi _{2}=(ka/2)\sin {(\theta )}.
\label{eq5}
\end{equation}%
represent the field delay per lattice period due to the finite propagation
speed. The limit of $\phi _{1,2}\rightarrow 0$ corresponds to the dense
lattice, with $ka\gg 1$.

The third term in Eq. (\ref{eq2}) corresponds to the inter-dot coupling
through the tunnelling. To interpret, it we introduce tunnelling coupling
factors $\xi _{p,q}^{(a,b)}$ which are step functions of the discrete
spatial coordinates, $p$ and $q$:%
\begin{equation}
\xi _{p,q}^{(a,b)}=\left\{
\begin{array}{ll}
\xi ^{(a,b)}, & p,q=-(N_{1,2}/2),...-1,0,1,...,N_{1,2}/2, \\
0, & \mbox{in all other cases}.%
\end{array}%
\right.   \label{step}
\end{equation}

The Hamiltonian of the tunnelling interaction is, in the tight-binding
approximation,
\begin{eqnarray}
\hat{H}_{T}=-\hbar \sum_{p,q}(\xi _{p+1,q}^{(a)}|a_{p,q}\rangle \langle
a_{p+1,q}|+\xi _{p,q}^{(a)}|a_{p,q}\rangle \langle a_{p-1,q}|+\xi
_{p,q+1}^{(a)}|a_{p,q}\rangle \langle a_{p,q+1}|+\xi
_{p,q}^{(a)}|a_{p,q}\rangle \langle a_{p,q-1}|)  \nonumber \\
-\hbar \sum_{p,q}(\xi _{p+1,q}^{(b)}|b_{p,q}\rangle \langle b_{p+1,q}|+\xi
_{p,q}^{(b)}|b_{p,q}\rangle \langle b_{p-1,q}|+\xi
_{p,q+1}^{(b)}|b_{p,q}\rangle \langle b_{p,q+1}|+\xi
_{p,q}^{(b)}|b_{p,q}\rangle \langle b_{p,q-1}|).  \label{eq7}
\end{eqnarray}%
Coefficients $\xi^{(a,b)}$ in Eq. (\ref{eq7}) are accounted for the
tunnelling coupling between adjacent QDs in the excited and ground states,
respectively. They are defined as scalars due to the isotropy of individual
QDs and the array. The step-like structure in Eq. (\ref{step}) corresponds
to the operator notation for the tunnelling in the finite-size array: the
number of the adjacent QDs coupled by the tunnelling reduces up to three for
sites at \textit{edges} of the array, and to two at \textit{corners}.

The last term in Hamiltonian (\ref{eq2})\ represents the local-field
effects, in the Hartree-Fock-Bogoliubov approximation \cite{21}, \cite{31}:
\begin{equation}
\Delta \hat{H}=\frac{4\pi }{V}N_{\alpha ,\beta }\mu _{\alpha }\mu _{\beta
}\sum_{p,q}\left( \hat{\sigma}_{p,q}^{-}\langle \hat{\sigma}%
_{p,q}^{+}\rangle +\hat{\sigma}_{p,q}^{+}\langle \hat{\sigma}%
_{p,q}^{-}\rangle \right) ,  \label{eq8}
\end{equation}%
where $\mu _{\alpha ,\beta }$ and $N_{\alpha ,\beta }$ are components of the
dipolar-moment vector and depolarization tensor of the single QD, $V$ is the
volume of the single QD, and angle brackets denote averaging of the
corresponding operator with respect to the given quantum state. The
summations over doubly repeated indices have been omitted in accordance with
the usual convention. The depolarization tensor depends both on the QD
configuration and quantum properties of electron-hole pairs, given by
\begin{equation}
N_{\alpha ,\beta }=\frac{V}{4\pi }\int_{V}\int_{V}|\xi (\mathbf{r}%
)|^{2}\left\vert \xi \left( \mathbf{r}^{\prime }\right) \right\vert
^{2}G_{\alpha ,\beta }\left( \mathbf{r}-\mathbf{r}^{\prime }\right) d^{3}%
\mathbf{r}\,d^{3}\mathbf{{r}^{\prime },}  \label{eq9}
\end{equation}%
where $\xi (\mathbf{r})$ is the envelope of the wave-function of
electron-hole pair, and $G_{\alpha ,\beta
}(\mathbf{r}-\mathbf{r}^{\prime })$ is the Green's tensor of the
Maxwell's equations in the quasi-static limit \cite{18}. The
present formulation may be applied to 1D and 2D setting alike
\cite{28}. Using the Hartree-Fock-Bogoliubov approximation,
local-field effects for the single QD, were investigated in the
strong coupling regime in Refs. \cite{21}, and \cite{31}, and
thereafter demonstrated experimentally \cite{1new}.

The temporal evolution of single-particle excitations is governed by the Schr%
\"{o}dinger equation,
\begin{equation}
i\hbar \frac{\partial |\Psi \rangle }{\partial t}=\hat{H}|\Psi \rangle .
\label{eq10}
\end{equation}%
The unknown wave function is taken in the form of a coherent superposition,
\begin{equation}
|\Psi \rangle =\sum_{p,q}(\Psi _{p,q}(t)\,e^{\left( i/2\right) \left( p\phi
_{1}+q\phi _{2}-\omega \,t\right) }|a_{p,q}\rangle +\Phi
_{p,q}(t)\,e^{-\left( i/2\right) \left( p\phi _{1}+q\phi _{2}-\omega
t\right) }|b_{p,q}\rangle ),  \label{eq11}
\end{equation}%
where $\Psi _{p,q}(t),\,\Phi _{p,q}(t)$ are probability amplitudes to be
found. Projection of the Schr\"{o}dinger equation onto the chosen basis
leads to the following system of coupled nonlinear evolution equations for
the probability amplitudes:
\begin{eqnarray}
\frac{\partial \Psi _{p,q}}{\partial t}&=&iF\Psi _{p,q}  \nonumber \\
&+& i\left( \xi _{p,q}^{(a)}\Psi _{p-1,q}e^{-i\varphi _{1}}+\xi
_{p+1,q}^{(a)}\Psi _{p+1,q}e^{i\varphi _{1}}+\xi _{p,q}^{(a)}\Psi
_{p,q-1}e^{-i\varphi _{2}}+\xi _{p,q+1}^{(a)}\Psi _{p,q+1}e^{i\varphi
_{2}}\right)  \nonumber \\
&-&ig\Phi _{p,q}-i\Delta \omega |\Phi _{p,q}|^{2}\Psi _{p,q},  \nonumber \\
\frac{\partial \Phi _{p,q}}{\partial t}&=&-iF\Phi _{p,q}  \nonumber \\
&+&i\left( \xi _{p,q}^{(b)}\Phi _{p-1,q}e^{i\varphi _{1}}+\xi
_{p+1,q}^{(b)}\Phi _{p+1,q}e^{-i\varphi _{1}}+\xi _{p,q}^{(b)}\Phi
_{p,q-1}e^{i\varphi _{2}}+\xi _{p,q+1}^{(b)}\Phi _{p,q+1}e^{-i\varphi
_{2}}\right)  \nonumber \\
&-&ig\Psi _{p,q}-i\Delta \omega |\Psi _{p,q}|^{2}\Phi _{p,q},  \label{eq12}
\end{eqnarray}%
where $F$ is the detuning parameter \cite{31}, $g\equiv -\mathbf{\mu E}%
_{0}/(2\hbar )$ is QD-field coupling factor, and $\Delta \omega \equiv 4\pi
\mu _{\alpha }\mu _{\beta }N_{\alpha ,\beta }/(\hbar V)$ is the
depolarization shift. The normalization condition for system (\ref{eq12}) is
\begin{equation}
\sum_{p=-N_{1}/2}^{N_{1}/2}\sum_{q=-N_{2}/2}^{N_{2}/2}(|\Psi
_{p,q}|^{2}+|\Phi _{p,q}|^{2})=1.  \label{eq13}
\end{equation}%
The observable value of the energy is given by
\begin{eqnarray}
\varepsilon &\equiv& \langle \hat{H}\rangle =\frac{1}{2}
\sum_{p=-N_{1}/2}^{N_{1}/2}\sum_{q=-N_{2}/2}^{N_{2}/2}\left[ -F(|\Psi
_{p,q}|^{2}-|\Phi _{p,q}|^{2})\right.  \nonumber \\
&-&\Phi _{p,q}^{\ast }\left( \xi _{p,q}^{(b)}\Phi _{p-1,q}e^{i\varphi
_{1}}+\xi _{p+1,q}^{(b)}\Phi _{p+1,q}e^{-i\varphi _{1}}+\xi _{p,q}^{(b)}\Phi
_{p,q-1}e^{i\varphi _{2}}+\xi _{p,q+1}^{(b)}\Phi _{p,q+1}e^{-i\varphi
_{2}}\right)  \nonumber \\
&-&\Psi _{p,q}^{\ast }\left( \xi _{p,q}^{(a)}\Psi _{p-1,q}e^{-i\varphi
_{1}}+\xi _{p+1,q}^{(a)}\Psi _{p+1,q}e^{i\varphi _{1}}+\xi _{p,q}^{(a)}\Psi
_{p,q-1}e^{-i\varphi _{2}}+\xi _{p,q+1}^{(a)}\Psi _{p,q+1}e^{i\varphi
_{2}}\right)  \nonumber \\
& & \left. +g\Psi _{p,q}^{\ast }\Phi _{p,q}+\frac{1}{2}\Delta \omega |\Psi
_{p,q}|^{2}|\Phi _{p,q}|^{2}+\mathrm{c.c.}\right] .  \label{eq14}
\end{eqnarray}%
The analysis of the Rabi solitons in the 2D QD lattices is developed below
on the basis of Eq. (\ref{eq12}).

In the simplest case, one may assume the inter-site coupling coefficients
for the ground and excited states to be equal, $\xi ^{(a)}=\xi ^{(b)}=\xi $.
Next, we set, by means of an obvious rescaling, $g=-1$ and $\mathrm{sign}%
(\Delta \omega )=-1$. These signs imply the attractive onsite linear
interaction between fields $\Psi _{p,q}$ and $\Phi _{p,q}$, and the
self-focusing sign of the XPM (cross-phase-modulation, i.e., the cubic
interaction between the different components) onsite nonlinearity. Actually,
if $g$ is originally positive, it can be made negative by substitution $\Phi
_{p,q}=-\tilde{\Phi}_{p,q}$. And if $\Delta \omega $ is originally positive,
it may be made negative by means of the usual staggering substitution \cite%
{1}. Thus, in Eqs. (\ref{eq12}) there remain three independent parameters
which can be combined into the frequency detuning, $F$, and the complex
lattice coupling, $\xi \exp {(i\phi _{1,2})}$. We here restrict $\phi _{1,2}$
to the basic interval, $0\leq \phi _{1,2}\leq \pi /2$, therefore both
positive and negative values of $F$ should be considered.

\subsection{The cylindrical-wave drive}

The spatial structure of Rabi waves is tunable by selecting the
corresponding driving field. Of particular interest is cylindrical drive,
defined as
\begin{equation}
\mathbf{E}(\mathbf{r},t)=\mathbf{e}_{z}\mathrm{Re}\left\{ \mathbf{E}%
_{0}\,H_{0}^{(1)}(k|\mathbf{r}-\mathbf{r}_{0}|)\exp {(-i\omega \,t)}\right\}
,  \label{eq15}
\end{equation}%
where $H_{0}^{(1)}(x)$ is the Hankel function of the zero order
and the first type. Such field may be excited by the infinitely
thin and infinitely long current wire placed at point
$\mathbf{r}_{0}$ normally to the array's plane (see Fig.
\ref{waves} (b)). The current may be produced, for example, by a
semiconductor quantum
wire excited by the exciton-polariton \cite{45}. The current point $\mathbf{r%
}_{0}$ is placed at the center of the cell consisting of dots with the
indices $(p=0,q=0),(p=-1,q=-1),(p=0,q=-1)$ and $(p=-1,q=0)$. The general
statement of the problem is similar to the plane-wave case, the difference
being only in the form of the $H_{df}$ component of Hamiltonian (\ref{eq1}),
which is, now,
\begin{equation}
\hat{H}_{df}=-\mathbf{\mu e}_{z}E_{0}\sum_{p,q}\hat{\sigma}%
_{p,q}^{+}S_{p,q}+H.c,  \label{eq16}
\end{equation}%
where
\begin{equation}
S_{pq}\equiv \frac{H_{0}^{(1)}\left( ka\sqrt{(p+1/2)^{2}+(q+1/2)^{2}}\right)
}{H_{0}^{(1)}\left( ka/\sqrt{2}\right) }.  \label{eq17}
\end{equation}

The corresponding equations of motion are
\begin{eqnarray}
\frac{\partial \Psi _{p,q}}{\partial t}=iF\Psi _{p,q}  \nonumber \\
+i\left[ \xi _{p,q}^{(a)}\Psi _{p-1,q}\varsigma _{pq}^{p-1,q}+\xi
_{p+1,q}^{(a)}\Psi _{p+1,q}\varsigma _{pq}^{p+1,q}+\xi _{p,q}^{(a)}\Psi
_{p,q-1}\varsigma _{pq}^{p,q-1}+\xi _{p,q+1}^{(a)}\Psi _{p,q+1}\varsigma
_{pq}^{p,q+1}\right]  \nonumber \\
-ig|S_{pq}|\Phi _{p,q}-i\Delta \omega |S_{pq}||\Phi _{p,q}|^{2}\Psi _{p,q},
\nonumber \\
\frac{\partial \Phi _{p,q}}{\partial t}=-iF\Phi _{p,q}  \nonumber \\
+i\left[ \xi _{p,q}^{(b)}\Phi _{p-1,q}\varsigma _{pq}^{*p-1,q}+\xi
_{p+1,q}^{(b)}\Phi _{p+1,q} \varsigma_{pq}^{*p+1,q} +\xi _{p,q}^{(b)}\Phi
_{p,q-1} \varsigma _{pq}^{*p,q-1}+\xi _{p,q+1}^{(b)}\Phi _{p,q+1} \varsigma
_{pq}^{*p,q+1} \right]  \nonumber \\
- ig|S_{pq}|\Psi _{p,q}-i\Delta \omega |S_{pq}||\Psi _{p,q}|^{2}\Phi _{p,q},
\label{eq18}
\end{eqnarray}

\begin{equation}
\varsigma _{pq}^{mn}\equiv \sqrt{\frac{S_{mn}}{S_{pq}}},  \label{eq19}
\end{equation}
supplemented by the normalization condition:
\begin{equation}
\sum_{p=-N_{1}/2}^{N_{1}/2}\sum_{q=-N_{2}/2}^{N_{2}/2}\left( |\Psi
_{p,q}|^{2}+|\Phi _{p,q}|^{2}\right) |S_{pq}|=1.  \label{eq20}
\end{equation}%
Actually, Eq. (\ref{eq18}) describes other types of driving field too.
Namely, substitution $k\rightarrow i\,k$ (for real $k$) transforms the
Hankel function into the correspondent Macdonald function $K_{0}(x)$, which
represents driving the Rabi wave by surface plasmons guided by a noble-metal
wire \cite{18}, or carbon nanotube \cite{46}.

\section{2D Rabi solutions}

\subsection{Dispersion relations}

In the limit of $\Delta \omega =0$, Eq. (\ref{eq12}) without the local-field
terms reduces to the 2D generalization of the Rabi-wave model considered in
Refs. \cite{26,28}. In this case, the dispersion relation can be derived by
looking for solutions as
\begin{equation}
\left\{ \Psi _{p,q},\Phi _{p,q}\right\} =\{A,B\}\exp \left( {i(\kappa
_{x}\,p+\kappa _{y}\,q)}\right) \exp {(-i\Omega \,t)},  \label{eq21}
\end{equation}%
where $\kappa _{x,y}$ are wavenumbers, $A,B$ are unknown wave amplitudes,
and $\Omega $ is an unknown frequency.

\begin{figure}[h]
\center\includegraphics[width=9cm]{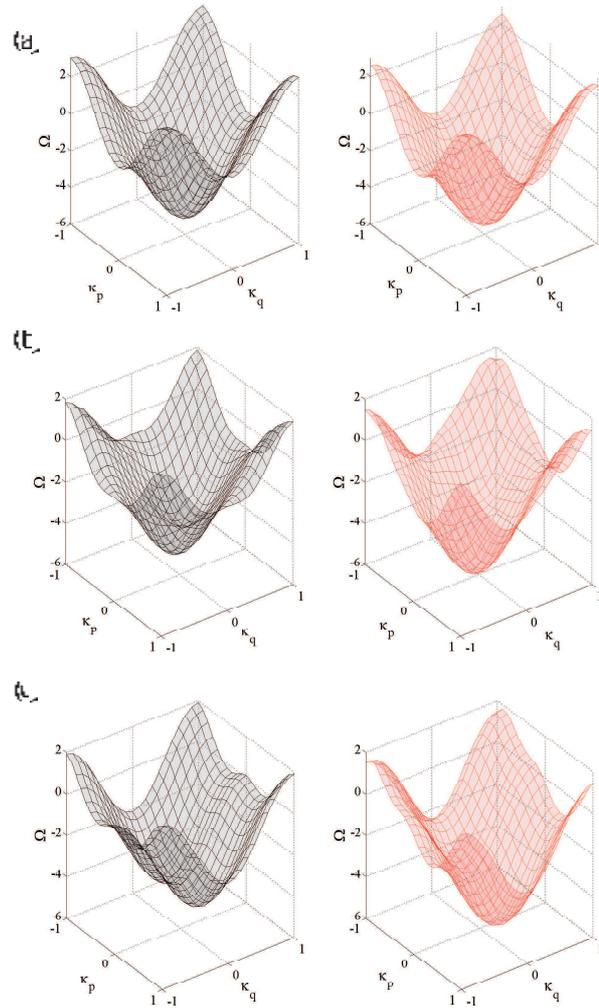}
\caption{(Color online) Dispersion curves $\Omega=\Omega _{2}(\protect\kappa %
_{p},\protect\kappa _{q})$ for the linearized system with the plane-wave
excitation: $F=0$ (black surfaces), $F=1$ (red surfaces), and $\protect\phi %
_{1}=\protect\phi _{2}=0$ (a), $\protect\phi _{1}=\protect\phi _{2}=\protect%
\pi /4$ (b), $\protect\phi _{1}=0,\,\protect\phi _{2}=\protect\sqrt{2}%
\protect\pi /4$ (c). Other coefficients are $\protect\xi =1$ and $g=-1$.
Discrete solitons are expected to exist in the semi-infinite gap. }
\label{dispersion}
\end{figure}

A straightforward analysis yields two branches of the dispersion relation,
\begin{eqnarray}
\Omega _{1,2}(\kappa _{x},\kappa _{y}) &=&-2\xi \left[ \cos {(\kappa _{x})}%
\cos {(\phi _{1})}+\cos {(\kappa _{y})}\cos {(\phi _{2})}\right]  \nonumber
\\
&\pm &\sqrt{g^{2}+\left\{ F-2\xi \left[ \sin {(\kappa _{x})}\sin {(\phi _{1})%
}+\sin {(\kappa _{y})}\sin {(\phi _{2})}\right] \right\} ^{2}},  \label{disp}
\end{eqnarray}%
examples of which are shown in Fig. \ref{dispersion} for $g=-1$ and $\xi =1$.

In the limit of $\phi _{1}=\phi _{2}=\phi =0$, Eq. (\ref{disp}) goes over
into the known dispersion relation for the usual system of linearly coupled
2D DNLS equations, cf. Ref. \cite{1}. It features two similar branches
shifted by a constant, $\Delta \Omega =2\sqrt{g^{2}+F^{2}}$, and corresponds
to the Rabi waves. In the case of $\phi _{1},\,\phi _{2}\neq 0$, shift $%
\Delta \Omega $ is no longer constant, as it depends on wavenumbers $\kappa
_{x}$ and $\kappa _{y}$.

Note that
\begin{equation}
\Omega (\kappa _{x},\kappa _{y},\phi _{1},\phi _{2})=\Omega (-\kappa
_{x},-\kappa _{y},-\phi _{1},-\phi _{2}),  \label{eqp1}
\end{equation}%
but
\begin{equation}
\Omega (\kappa _{x},\kappa _{y},\phi _{1},\phi _{2})\neq \Omega (-\kappa
_{x},-\kappa _{y},\phi _{1},\phi _{2}).  \label{eqp2}
\end{equation}%
Equation (\ref{eqp1}) shows that the inversion of the propagation direction
of the driving field inverts the direction of the Rabi-wave propagation.
This is according to the non-reciprocity of Rabi-waves in 1D chains, noted
in Ref. \cite{43}.

\subsection{Constructing Rabi solitons and their radiation patterns}

Equations (\ref{eq12}) may give rise to stationary discrete 2D fundamental
solitons and vortices, looking for localized solutions as
\begin{equation}
\Psi _{p,q}=e^{-i\Omega t}A_{p,q},~\Phi _{p,q}=e^{-i\Omega t}B_{p,q},
\label{rabisol}
\end{equation}%
where $\Omega $ is the corresponding carrier frequency, and $%
A_{p,q},\,B_{p,q}$ are localized complex lattice fields vanishing at
infinity (in terms of the numerical solutions, they vanish at boundaries of
the computation domain). Stationary soliton solutions of Eqs. (\ref{eq12})
were numerically obtained by adopting the nonlinear equation solver based on
the Powell method while direct dynamical simulations were based on the
Runge-Kutta procedure of the sixth order, \cite{47}, \cite{48}. The
numerical solution was constructed for a finite lattice of size $21\times 21$
(unless stated otherwise), with periodic boundary conditions. We have
checked that a larger lattice produces the same results.

The emitted field is described by the field operator in the Heisenberg
representation \cite{19},
\begin{equation}
\hat{\mathbf{E}}(\mathbf{r},t)=-\frac{1}{4\pi \epsilon _{0}}(\nabla \cdot
\nabla -\frac{1}{c^{2}}\frac{\partial ^{2}}{\partial t^{2}})\int_{V}\frac{1}{%
|\mathbf{r}-\mathbf{r}^{\prime }|}\hat{\mathbf{P}}(t-\frac{|\mathbf{r}-%
\mathbf{r}^{\prime }|}{c})d^{3}\mathbf{r}^{\prime }.  \label{eq22}
\end{equation}%
Here, operator $\hat{\mathbf{P}}$ of the induced polarization describes the
displacement current induced by the soliton's profile, which may be
considered as an external source, from point of view of the antenna. The
onsite polarization operator, written in terms of creation-annihilation
operators is
\begin{equation}
\hat{\mathbf{P}}_{p,q}(t)=\mathbf{\mu }\hat{\sigma}_{p,q}^{+}(t)+\mathrm{H.c.%
}  \label{eq23}
\end{equation}%
Due to the smallness of the field produced by the soliton in the QD array,
its emission is determined by the linear approximation, hence the total
field is built as a superposition of partial fields emitted by different QDs
independently. The plane-wave drive is, thus,
\begin{eqnarray}
\hat{\mathbf{E}}(\mathbf{r},t) &=&-\frac{V}{4\pi \epsilon _{0}}\left( \nabla
\cdot \nabla -\frac{1}{c^{2}}\frac{\partial ^{2}}{\partial t^{2}}\right)
\mathbf{e}_{z}  \nonumber \\
&&\times \sum_{p=-N_{1}/2}^{N_{1}/2}\sum_{q=-N_{2}/2}^{N_{2}/2}\frac{\hat{P}%
_{zp,q}\left( t-c^{-1}\left\vert \mathbf{r}-\mathbf{e}_{x}pa-\mathbf{e}%
_{y}qa\right\vert \right) }{\left\vert \mathbf{r}-\mathbf{e}_{x}pa-\mathbf{e}%
_{y}qa\right\vert }.  \label{eq24}
\end{eqnarray}%
To obtain the observable value of field, it is necessary to replace the
polarization operator by its main value with respect to the quantum state (%
\ref{eq11}), $\langle \Psi |\hat{P}_{zp,q}|\Psi \rangle $. The result is
\begin{equation}
P_{zp,q}(t)=\langle \hat{P}_{sp,q}(t)\rangle =\frac{\mu }{V}\Psi
_{p,q}(t)\Phi _{p,q}^{\ast }\exp \left[ {i(p\phi _{1}+q\phi _{2})}\right]
\exp {(-i\omega \,t)}+\mathrm{c.c.}  \label{eq25}
\end{equation}%
It is convenient to present the far-zone field in the spherical coordinates
with the origin set at the central point of the cell with $p=q=0$, defined
as $x=R\sin {(\theta )}\,\cos {(\varphi )},\,y=R\sin {(\theta )}\,\sin {%
(\varphi )},\,z=R\cos {(\theta )}$. The terms of order $O(R^{-2})$ should be
omitted, and the radial (longitudinal) component of the electric field
vanishes too. As a result, we obtain, for the quasi-spherical observable
field:
\begin{equation}
\mathbf{E}_{\mathrm{Rad}}=\lim_{R\rightarrow \infty }\mathbf{{E}=}\frac{\mu
\mathbf{e}_{\theta }}{4\pi \epsilon _{0}}\frac{\omega ^{2}}{c^{2}}\frac{%
e^{-i\omega (t-R/c)}}{R}F\left( \theta ,\varphi ,\phi _{1},\phi _{2};t-\frac{%
R}{c}\right) +\mathrm{c.c.}\mathbf{,}  \label{eq26}
\end{equation}%
with the radiation pattern
\begin{eqnarray}
F(\theta ,\varphi ,\phi _{1},\phi _{2};t-\frac{R}{c}) =\sin {\theta }%
\sum_{p=-N_{1}/2}^{N_{1}/2}\sum_{q=-N_{2}/2}^{N_{2}/2}\Psi _{p,q}(\tilde{t}%
)\Phi _{p,q}^{\ast }(\tilde{t})  \nonumber \\
\times \exp \{ip\left[ \frac{\omega \,a}{c}\sin {\theta }\cos {\varphi }%
+2\phi _{1}\right] +iq\left[ \frac{\omega \,a}{c}\sin {\theta
}\sin {\varphi }+2\phi _{2}\right] \},  \label{antplanar}
\end{eqnarray}%
where $\tilde{t}\equiv t-R/c+c^{-1}\sin {\theta }\left[ p\cos {\varphi }%
+q\,\sin {\varphi }\right] $. We stress that, in contrast with classical
macroscopic antennas, the radiation pattern given by Eq. (\ref{antplanar})
is non-steady, as it exhibits a slow dependence on the spatiotemporal
variable, $t-R/c$, which corresponds to the amplitude and frequency
modulation of the field due to the time dependence of probability amplitudes
in quantum state (\ref{eq11}). The stable soliton produces, through Eq. (\ref%
{antplanar}), a radiation pattern in the form of
\begin{eqnarray}
F(\theta ,\varphi ,\phi _{1},\phi _{2}) &=&\sin {\theta }%
\sum_{p=-N_{1}/2}^{N_{1}/2}\sum_{q=-N_{2}/2}^{N_{2}/2}A_{p,q}B_{p,q}^{\ast
}\exp \left\{ {ip}\left[ \frac{\omega \,a}{c}\sin {\theta }\cos {\varphi }{%
+2\phi _{1}}\right] \right\}   \nonumber \\
&&\times \exp \left\{ {iq}\left[ \frac{\omega \,a}{c}\sin {\theta }\sin {%
\varphi }+2\phi _{2}\right] \right\} .  \label{antplanar11}
\end{eqnarray}%
In particular, the radiation pattern produced by the 1D in the 2D space can
be written in a simple form by setting $N_{2}=0$ in Eq. (\ref{antplanar11}):
\begin{equation}
F(\theta ,\varphi ,\phi )=\sin {(\theta )}\sum_{p=-N/2}^{N/2}A_{p}B_{p}^{%
\ast }\exp \left\{ {ip}\left[ \frac{\omega \,a}{c}\left( {\sin {\theta }}%
\right) \left( {\cos {\varphi }}\right) {+2\phi }\right] \right\} ,
\label{antplanar2}
\end{equation}%
where the notation is changed as $A_{p,q}\rightarrow
A_{p},\,B_{p,q}\rightarrow B_{p},\,\phi _{1}\rightarrow \phi
,\,N_{1}\rightarrow N$, to comply with that adopted in Ref. \cite{31}.

It is relevant to note that, according to Ref. \cite{43} the radiation
pattern is subject to the Onsager kinetic relations, which take into regard
symmetry constraints \cite{51}. As it follows from Eqs. (\ref{eq12}) and (%
\ref{rabisol}), $(A_{-p,-q}B_{-p,-q}^{\ast })=(A_{p,q}B_{p,q}^{\ast })$,
hence
\begin{equation}
F(\pi -\theta ,\varphi ,-\phi _{1},-\phi _{2})=F(\theta ,\varphi ,\phi
_{1},\phi _{2}),  \label{kk}
\end{equation}%
while $F(\pi -\theta ,\varphi ,\phi _{1},\phi _{2})\ne F(\theta ,\varphi
,\phi _{1},\phi _{2})$. It shows that, rotating the QD array by $180^{%
\mathrm{o}}$, one should invert the propagation direction of the driving
field to keep the physical state invariant, similar to the the sign of the
angular velocity in the rotating liquid, or the sign of the magnetic field
\cite{51}. This symmetry agrees with the non-reciprocity of the Rabi waves
exhibited by Eq. (\ref{eqp1}).

For the cylindrical drive, the radiation pattern in the far-field zone is
obtained from Eq. (\ref{eq26}) in a similar way:
\begin{equation}
F(\theta ,\varphi )=\sin {\theta }\sum_{p=-N_{1}/2}^{N_{1}/2}%
\sum_{q=-N_{2}/2}^{N_{2}/2}A_{p,q}B_{p,q}^{\ast }|S_{p,q}|\exp \left\{\frac{%
i\omega \, a}{c}\sin {\theta } \left[ p\cos{\varphi} +q\sin {\varphi} \right]
\right\} .  \label{antcyl1}
\end{equation}

The useful information in antenna context are the radiation patterns in the $%
E$- and $H$ -planes defined, respectively, as $F_{e}(\theta )\equiv F(\theta
,\varphi =\pi /2)$ and $F_{h}(\theta )\equiv F(\theta =\pi /2,\varphi )$
\cite{32}. Plots of these quantities follow the corresponding dynamically
stable soliton evolution in figures below.

\section{Results and discussion}

\subsection{Rabi solitons driven by the plane wave}

After stationary solitons were found as outlined above [see Eq. (\ref%
{rabisol})], their stability against small perturbations was examined in the
framework of linearized equations for small perturbations added to the
stationary solution,
\begin{equation}
\Psi _{p,q}(t)=\left[ A_{p,q}+\delta A_{p,q}(t)\right] e^{-i\Omega t},\,\Phi
_{p,q}(t)=\left[ B_{p,q}+\delta B_{p,q}(t)\right] e^{-i\Omega t}.
\label{perturbations}
\end{equation}%
In this case, the linearization of Eq. (\ref{eq12}) with respect to the
perturbations $\delta A_{p,q}$ and $\delta B_{p,q}$, leads to a linear
eigenvalue (EV) problem for the instability growth rate of small
perturbations. Obviously, the soliton is stable if all the eigenvalues have
zero real parts. On the other hand, the presence of EVs with a positive real
part indicates at exponential or oscillatory instability, in the cases of
purely real or complex EVs, respectively. Finally, the dynamical stability
of the discrete solitons was verified by dint of direct simulations of Eqs. (%
\ref{eq12}).

It is well known that the single-component 2D DNLS equation gives
rise to three types of fundamental discrete solitons: stable
onsite, and unstable hybrid and inter-site ones \cite{49}, along
with several types of vortices with vorticity $S=1$ (eight
configurations) and $S=2$ (four configurations)
\cite{49,50,vortex}. The vorticity is identified as the winding
number of the phase of the lattice field, i.e., the total change
of the phase along a closed curve surrounding the pivotal point of
the vortex, divided by $2\pi $. We here consider the values of
$S=1$ and $S=2$.

\begin{figure}[th]
\center\includegraphics [width=12cm]{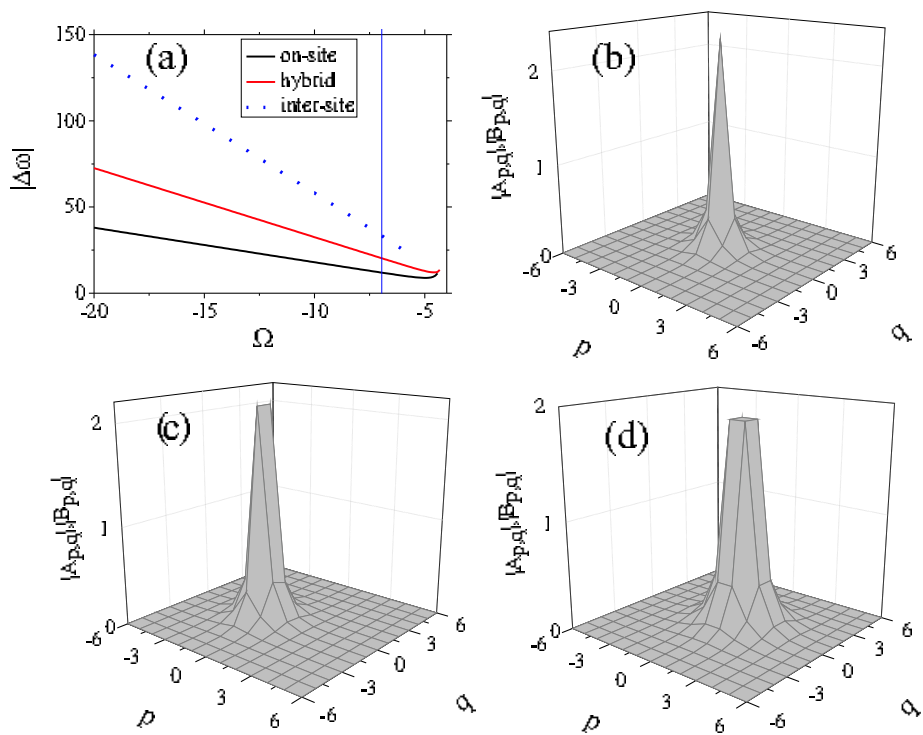}
\caption{Families of fundamental solitons, formed by two identical onsite
(solid black line), hybrid (solid red line) and offsite (dotted green line)
components, are represented by the corresponding $|\Delta \protect\omega %
(\Omega )|$ dependencies in (a). Examples of these solitons for $\Omega =-7$
[the vertical blue line in (a) which intersects all three curves] are shown
in (b) -- onsite, (c) -- hybrid and (d) -- offsite, respectively. This
figure pertains to the model with the plane-wave excitation, see Eq. (%
\protect\ref{eq12}). }
\label{slika1}
\end{figure}

Obviously, the soliton complexes must be located inside the semi-infinite
gap of dispersion relation (\ref{disp}), see Fig. \ref{dispersion}. The most
straightforward counterpart of the single-component 2D DNLS equation is the
model of the QD array based on Eq. (\ref{eq12}) with zero detuning, $F=0$
and $\phi _{1}=\phi _{2}=0$. In this system, we find soliton complexes
formed by two identical real components of the onsite, hybrid, and
inter-site types (Fig. \ref{slika1}), and families of discrete vortex
solitons with topological charges $S=1$ and $S=2$ (together with
quadrupoles), which are supported by the attractive XPM interaction. The
localized modes in each component are similar to the corresponding types of
fundamental and vortex solitons formed in the single-component 2D DNLS
equation.

\begin{figure}[h]
\center\includegraphics [width=12cm]{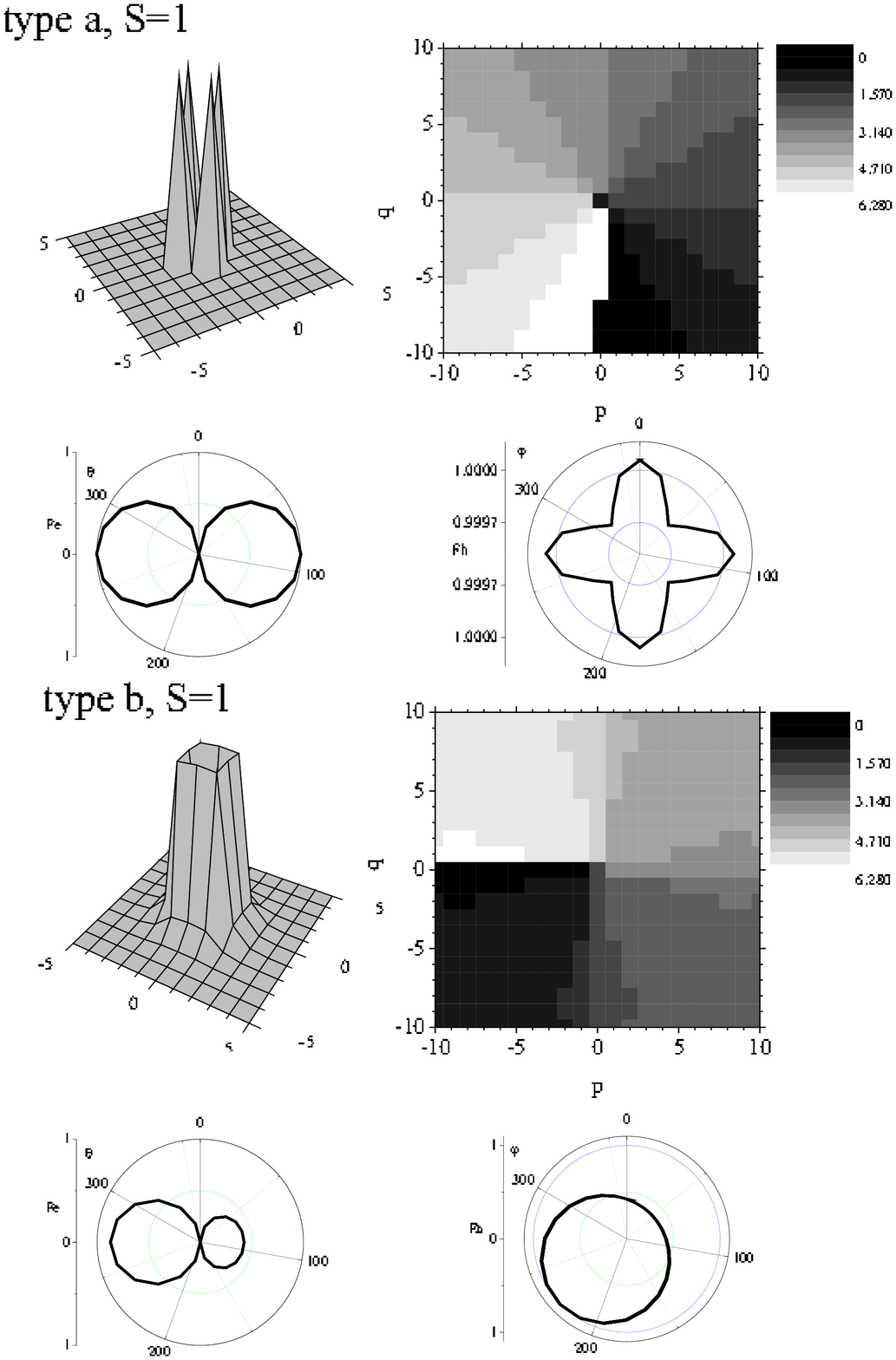}
\caption{Amplitude, phase profiles, and radiative patterns in $E$- and $H$%
-planes of vortices with $S=1$ defined as types $a$ and $b$ in Ref.
\protect\cite{vortex}. The parameters in Eq. (\protect\ref{eq12}) are $%
\protect\xi =0.01,\,F=0,\,\protect\phi _{1}=\protect\phi _{2}=0$, and$%
\,\Omega =-19.3$. Both are stable according to the linear-stability analysis
and dynamical simulations.}
\label{vortex1}
\end{figure}

Different types of vortices in each component are schematically shown in
tables I and II. Configurations $a$, $c$, $d$, $e$, $f$, $g$, $i$, and $j$
represent possible structures of onsite vortices, with the pivotal point
located on a lattice site, while $b$, $h$, $k$, and $l$ generate off-site
vortices, whose pivot falls located between lattice sites (the same
nomenclature as in Ref. \cite{50}). An example of the vortex of type $a$ is
plotted in Fig. \ref{vortex1}.

\begin{figure}[th]
\center\includegraphics [width=12cm]{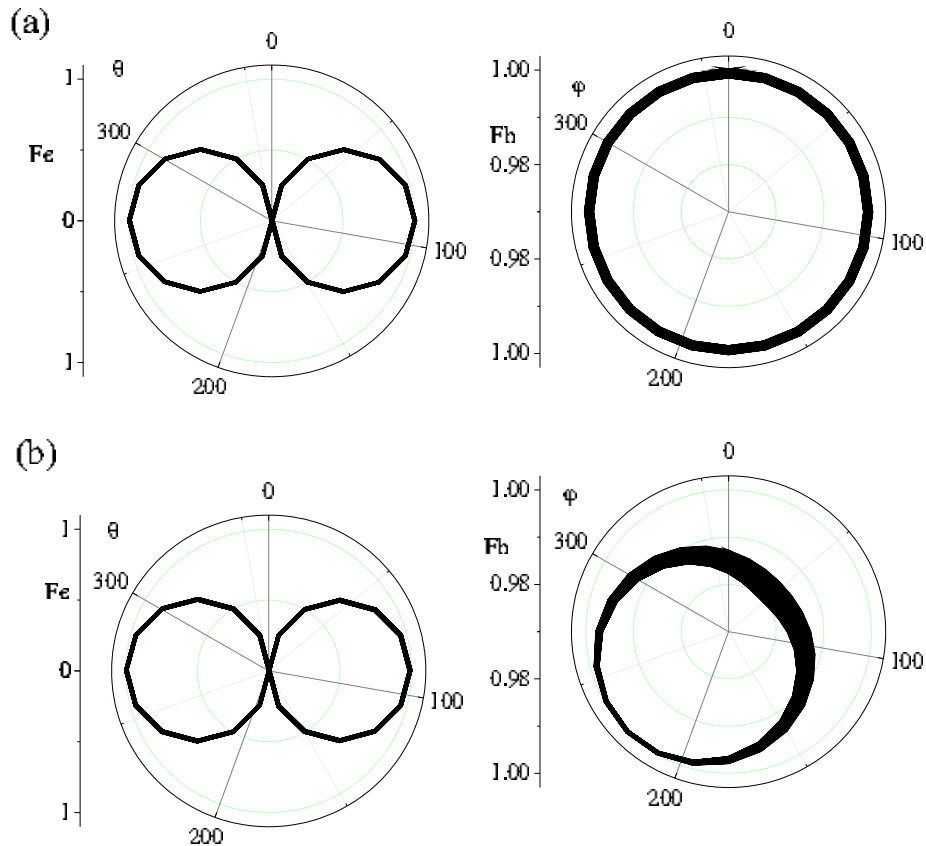}
\caption{The radiative patterns in the $E$- and $H$-directions, $F_{e}(%
\protect\theta )$ and $F_{h}(\protect\phi )$, for the complex consisting of
two fundamental onsite solitons with (a) $\Omega =-7,\,F=0,\,\protect\phi %
_{1}=\protect\phi _{2}=0$ and (b) $\Omega =-7,\,F=0,\,\protect\phi _{1}=%
\protect\phi _{2}=\protect\pi /4$. The plane-wave excitation is considered,
see Eq. (\protect\ref{eq12}). }
\label{slika1r}
\end{figure}

The linear-stability analysis shows that the complexes formed by identical
fundamental-soliton onsite components are stable, and that there are
stability regions in the respective parameter space for vortex
configurations of types $a$, $b$ (ground-state vortices), $e$, $g$ with $S=1$
(excited state vortices), as well as for types $j$, $k$ with $S=2$. The
corresponding radiation-field distributions for the onsite fundamental
solitons are plotted in Fig. \ref{slika1r}, while the same for the stable
vortices of the $a$ and $b$ types is shown in Fig. \ref{vortex1}. The
vortices of type $k$ actually feature an offsite quadrupole configuration
(rather than a truly vortical one), based on the four-site frame and sharp
phase changes in steps of $\pi$ between lattice sites. The stability regions
of all the above-mentioned complexes almost overlap with the stability
regions of their counterparts in the single-component 2D DNLS equation.
Briefly speaking, the vortex-stability range is narrow, particularly for
spatially broad vortices, and shrinks with the increase of $\xi $ (which is
the counterpart of the inter-site coupling constant $C$ in single-component
DNLS equation \cite{50}). This is in accordance with the fact that large $%
\xi $ corresponds to the continuum limit, in which all the localized vortex
modes are unstable. We have also checked a possibility to form vortex
complexes formed of two identical vortices with opposite signs (phase shift $%
\pi $), but they all turn out to be unstable.

\begin{figure}[h]
\center\includegraphics [width=12cm]{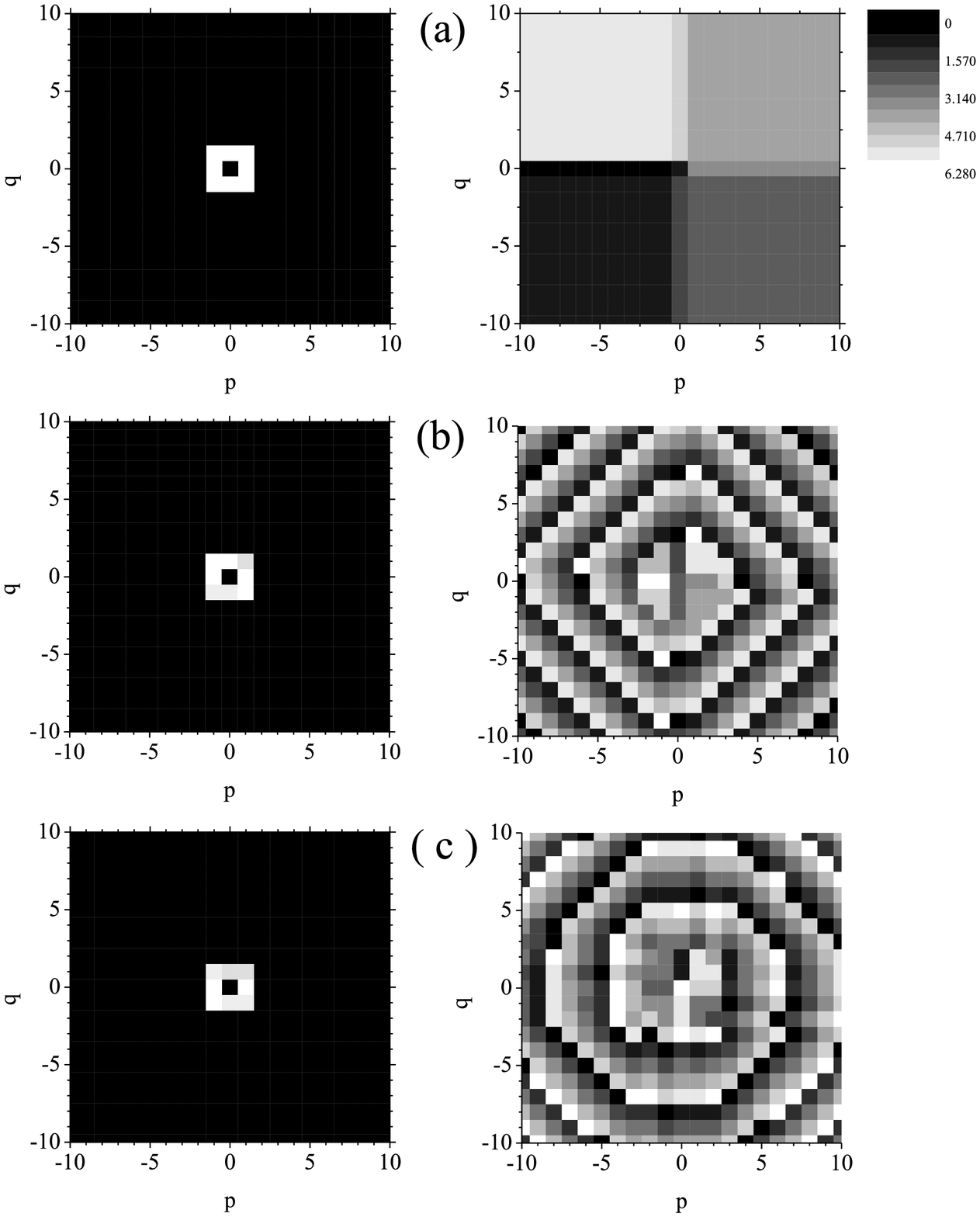}
\caption{The evolution of one component of an unstable vortex of type $c$
with $S=1$. (a) The amplitude and phase profiles of the stationary vortex
with $\Omega =-10$. Panels (b) and (c) show the profiles in the course of
the perturbed evolution ($t=50$ and $t=100$ in arbitrary units,
respectively). Other parameters are $\protect\xi =0.01,\ F=0,\,\protect\phi %
_{1}=\protect\phi _{2}=0$. }
\label{vortex3}
\end{figure}

The predicted dynamical stability and instability has been confirmed by
direct simulations for the onsite fundamental soliton complexes and certain
vortex complexes for typical values of the system's parameters. The onsite
fundamental complexes and those vortex complexes which were predicted to be
stable indeed keep their amplitude and phase structure (i.e., the vorticity)
in the course of the evolution. On the other hand, hybrid and offsite
complexes, whose instability is predicted by the linear-stability analysis,
radiate away a significant part of their norm and rearrange into dynamically
robust onsite-centered localized breathers. Direct simulations of vortex
complexes, which were predicted to be unstable, demonstrate that their
amplitude profiles are quite robust, and they keep the vorticity in the
course of the evolution. However, the symmetry with respect to the pivot is
broken, and the symmetry between their two components is broken too. The so
established mode may be categorized as an irregularly oscillating vortex
breather, see Fig. \ref{vortex3}.

In the 2D array of QDs with nonzero detuning ($F=1$ is chosen below, as a
characteristic example), all types of the above-mentioned fundamental and
vortical complexes can be generated, but the symmetry between the two
components ($A_{p,q}$ and $B_{p,q}$) is not preserved. Simultaneously, the
similarity to the single-component DNLS equation is lost. Namely, the
overall shape of the components in fields $A_{p,q}$ and $B_{p,q}$ is kept,
but amplitudes and widths of the components change, and their precise
symmetry about the central point is broken (slightly violated). In this
case, the linear-stability analysis does not predict the existence of
genuinely stable fundamental complexes, with the exception of the vortex
complexes of type $a$, with $S=1$, and of type $k$, with $S=2$, see below.
However, dynamical simulations demonstrate that localized modes of the
onsite type form robust breathing complexes, which keep the initial overall
structure (in particular, the vorticity is kept).

On the other hand, the computation of the eigenvalues for small
perturbations shows stability of vortex complexes of type $a$, with $S=1$,
as well as of types $k$ with $S=2$ in the presence of $F\neq 0$. Dynamical
simulations confirm their stability in the corresponding parameter regimes.
Other families of vortex complexes formed at $F\neq 0$ are unstable
according to both the linear-stability analysis and dynamical simulations.
In fact, the instability destroys their phase structure, while the amplitude
patterns remain quite robust.

Thus, the general conclusion concerning the system (\ref{eq12}) driven by
the plane wave \ is that, in the absence of the phase shifts, $\phi
_{1}=\phi _{2}=0$, the the stability of the complexes is similar to what is
known about their counterparts in the single 2D DNLS equation single 2D
lattice case. Namely, irrespective of the presence of the detuning, $F$, in
Eq. (\ref{eq12}), the those modes which were stable in the single-component
model, remain completely or effectively stable in the two-component one,
evolving in some cases from the stationary shapes into breathing ones, but
keeping the overall structure.

\begin{figure}[h]
\center\includegraphics [width=10cm]{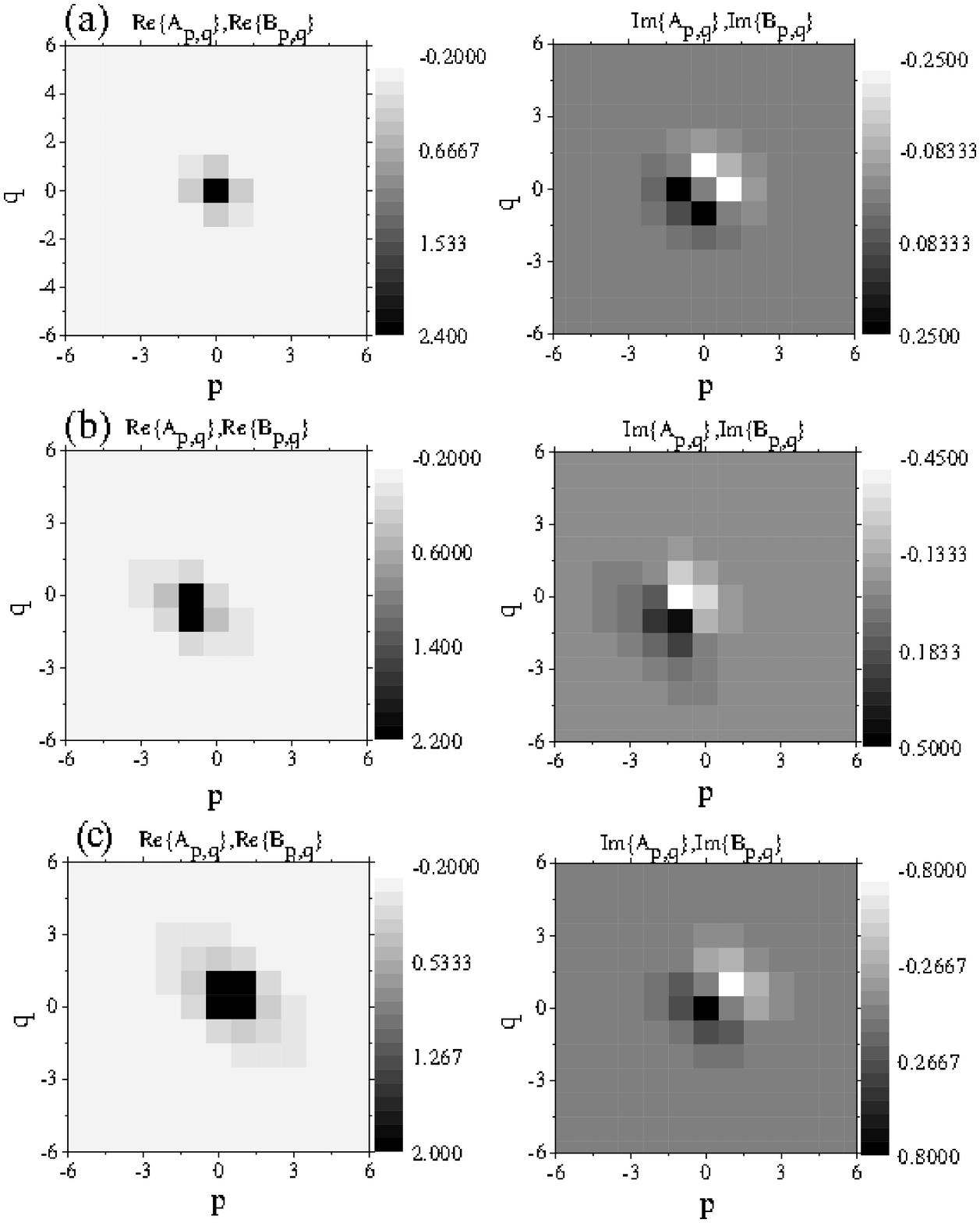}
\caption{ Examples of fundamental-soliton complexes [in the model with the
plane-wave excitation, Eq. (\protect\ref{eq12})], formed by identical
components, at $F=0$, $\Omega =-7$, and $\protect\phi _{1}=\protect\phi _{2}=%
\protect\pi /4$. Plotted are real and imaginary parts of the stationary
lattice fields for the complexes of the onsite (a), hybrid (b), and
intersite (c) types, respectively. }
\label{slika4}
\end{figure}

Considering the system with non-zero phase shifts, $\phi _{1}$ and $\phi
_{2} $ in Eq. (\ref{eq12}), we mainly focus on two configurations, \textit{%
viz}., the diagonal and anisotropic ones, with $\phi _{1}=\phi _{2}=\pi /4$
or $\phi _{1}=0,\,\phi _{2}=\sqrt{2}\pi /4$, respectively. In both these
cases, we have found complexes of all the above-mentioned types. Namely,
fundamental solitons of the onsite, hybrid and inter-site types, with
symmetric components, in the case of $F=0$ (see Fig. \ref{slika4}), and
their asymmetric counterparts at $F\neq 0$. The existence boundary for the
fundamental soliton complexes is slightly altered by the nonvanishing phase
shifts, as can be seen from small changes of the respective dispersion
surfaces displayed in Fig. \ref{dispersion}. Example of the soliton
radiation fields for a stable configuration is shown in Fig. \ref{slika1r}%
(b).

\begin{figure}[h]
\center\includegraphics [width=12cm]{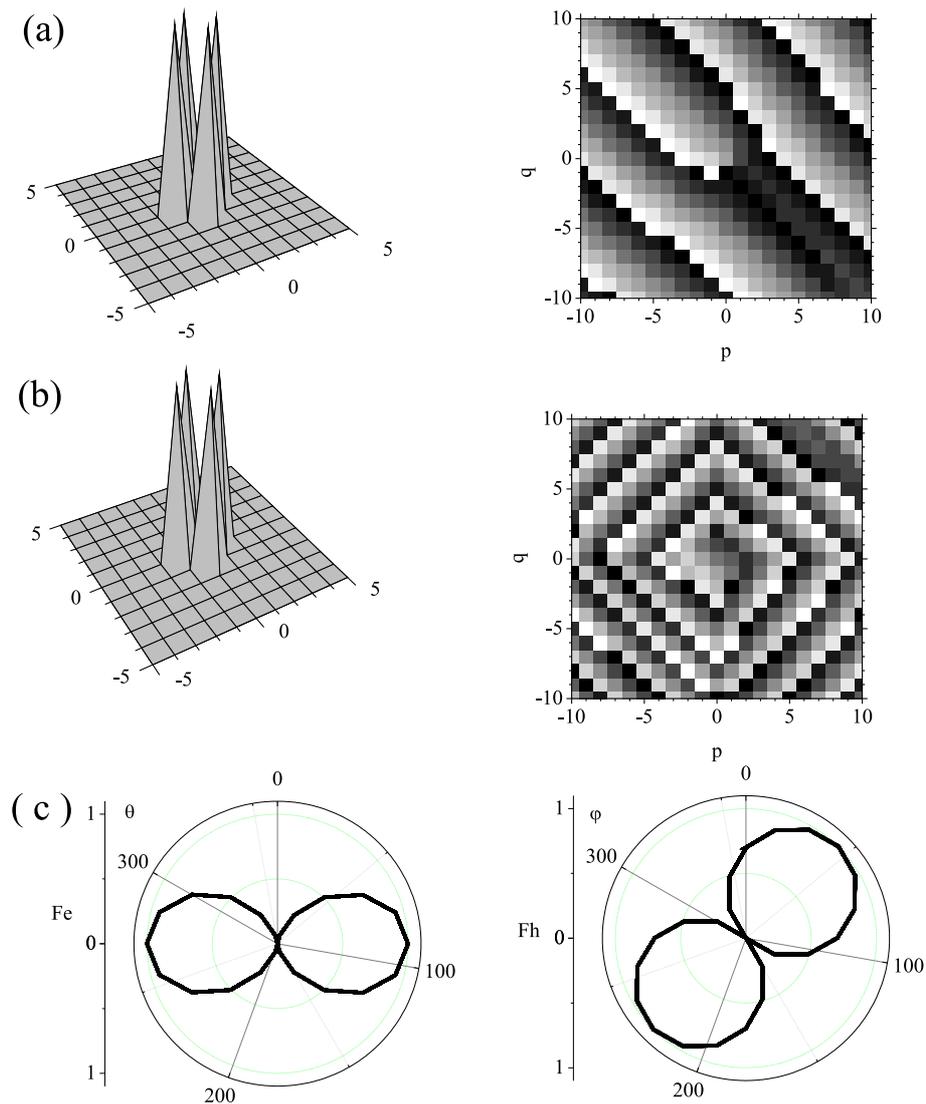}
\caption{Initial (a) amplitude and phase profiles of the vortex of type $a$,
in terms of Ref. \protect\cite{vortex}, the result of its evolution at $%
t=100 $ (b), and the respective radiative patterns $F_{e}$ and $F_{h}$ (c).
The parameters in Eq. (\protect\ref{eq12}) are $\protect\xi =0.01,\,F=0,\,%
\protect\phi _{1}=\protect\phi _{2}=\protect\pi /4$, and $\,\Omega =-18$.}
\label{vortex4}
\end{figure}

Counterparts of all the vortex field configurations, which are identified
above in the system with $\phi _{1,2}=0$, are also found in the system with $%
\phi _{1,2}\neq 0$. Differences are observed in the boundary of the
existence region, and parameter regions where robust vortex breathers
appear. In Fig. \ref{vortex4}, the evolution of the symmetric vortex soliton
of type $a$ is displayed for $\phi _{1}=\phi _{2}=\pi /4,\,F=0$. According
to the computation of perturbation eigenvalues, stable complexes of two $a$
type vortices (with topological charge $S=1$) keep their stability at $\phi
_{1}=\phi _{2}\neq 0$, while unequal phase shifts destabilize the complexes.
The radiation pattern keeps its shape, as shown in Fig. \ref{vortex4}(c).
Most robust are complexes formed by the quadrupoles of type $k$ with $S=2$,
in the sense that they are stable in a certain parameter area for all
relations between phases $\phi _{1}$ and $\phi _{2}$. Lastly, the phase
differences always destabilize other types of vortex complexes. In general,
the value of the corresponding instability rate, which is proportional to a
purely real eigenvalue, increases with increasing $\phi _{1},\,\phi _{2}$
and fixed $F$. Direct simulations confirm that only complexes formed by the
onsite fundamental solitons, and certain types of vortices are stable in
particular parameter ranges, chiefly in the form of robust localized
breathing patterns. Similar to what is mentioned above, the instability
destroys the phase structure of vortices, without essentially affecting the
amplitude pattern.

\begin{table}[tbp]
\caption{Schemes of the amplitude and phase profiles corresponding to the
two identical components of the discrete vortex with $S=1$ in the system
with $F=0,\protect\phi _{1}=\protect\phi _{2}=0$ [the model with the
plane-wave excitation, Eq. (\protect\ref{eq12})]. Explicitly written shares
of $\protect\pi $ (or $0$) are values of the phase at main sites carrying
the vortex complex. Symbol \textrm{x} designates sites with zero amplitude.
Only complexes with the components of the $b$ and $l$ types for $S=1$, and
of the $k$ and $h$ types for $S=2$, respectively, are found in the model
with the cylindrical-wave excitation. }
\label{tabela1}
\par
\begin{tabular}{|l|l|l|l|}
\hline
Type & $S$ & Phase configuration & Stable (plane wave/ cylindrical wave) \\
\hline
\ \ \ \ $a$ & \ \ 1 & \ \ \ \ \ \ \ \ \ \ \ \ \ \ \ 0\  & \ \ \ \ Yes/No \\
&  & \ \ \ \ \ \ \ \ $\pi /2$\ \ x\ \ $3\pi /2$ &  \\
&  & \ \ \ \ \ \ \ \ \ \ \ \ \ \ \ $\pi $\  &  \\ \hline
\ \ \ \ $b$ & \ \ 1 & \ \ \ \ \ \ \ \ \ \ \ \ 0 \ \ $3\pi /2$ & \ \ \ \
Yes/No \\
&  & \ \ \ \ \ \ \ \ \ \ \ \ $\pi /2$ \ \ $\pi $ &  \\ \hline
\ \ \ \ $c$ & \ \ 1 & \ \ \ \ \ \ \ \ $\pi /4$ \ \ \ 0 \ \ $7\pi /4$ & \ \ \
\ No/No \\
&  & \ \ \ \ \ \ \ \ $\pi /2$ \ \ \ x \ \ $3\pi /2$ &  \\
&  & \ \ \ \ \ \ \ $3\pi /4$ \ \ $\pi $ \ \ $5\pi /4$ &  \\ \hline
\ \ \ \ $d$ & \ \ 1 & \ \ \ \ \ \ \ \ \ \ \ \ \ \ \ \ 0 & \ \ \ \ No/No \\
&  & \ \ \ \ \ \ \ \ $\pi /4$ \ \ x \ $7\pi /4$ &  \\
&  & \ $\pi /2$ \ \ x \ \ \ \ x \ \ \ \ x \ \ $3\pi /2$ &  \\
&  & \ \ \ \ \ \ \ $3\pi /4$ \ x \ \ $5\pi /4$ &  \\
&  & \ \ \ \ \ \ \ \ \ \ \ \ \ \ \ $\pi $ &  \\ \hline
\ \ \ \ $e$ & \ \ 1 & \ \ \ \ \ \ \ \ \ \ \ \ \ \ \ \ 0 & \ \ \ \ Yes/No \\
&  & \ \ \ \ \ \ \ \ $\pi /4$ \ \ x \ $7\pi /4$ &  \\
&  & \ $\pi /2$ \ \ x \ \ \ \ 0 \ \ \ \ x \ \ $3\pi /2$ &  \\
&  & \ \ \ \ \ \ $3\pi /4$ \ \ x \ $5\pi /4$ &  \\
&  & \ \ \ \ \ \ \ \ \ \ \ \ \ \ \ $\pi $ &  \\ \hline
\ \ \ \ $f$ & \ \ 1 & \ \ \ \ \ \ \ \ \ \ \ \ \ \ \ \ 0 & \ \ \ \ No/No \\
&  & \ \ \ \ \ \ \ \ $\pi /4$ \ \ 0 \ \ $7\pi /4$ &  \\
&  & $\pi /2$ \ \ $\pi /2$ \ \ x \ \ $3\pi /2$ \ $3\pi /2$ &  \\
&  & \ \ \ \ \ \ $3\pi /4$ \ \ $\pi $ \ \ $5\pi /4$ &  \\
&  & \ \ \ \ \ \ \ \ \ \ \ \ \ \ \ $\pi $ &  \\ \hline
\ \ \ \ $g$ & \ \ 1 & \ \ \ \ \ \ \ \ \ \ \ \ \ \ \ 0 & \ \ \ \ Yes/No \\
&  & \ \ \ \ \ \ \ \ $\pi /4$ \ $\pi $ \ $7\pi /4$ &  \\
&  & $\pi /2$ \ $3\pi /2$ \ x \ \ $\pi /2$ \ $3\pi /2$ &  \\
&  & \ \ \ \ \ \ \ $3\pi /4$ \ 0 \ $5\pi /4$ &  \\
&  & \ \ \ \ \ \ \ \ \ \ \ \ \ \ \ $\pi $ &  \\ \hline
\ \ \ \ $h$ & \ \ 1 & \ \ \ \ \ \ \ \ \ \ \ 0 \ \ \ $7\pi /4$ & \ \ \ \ No/No
\\
&  & \ \ $\pi /4$ \ \ \ 0 \ \ \ $3\pi /2$ \ $3\pi /2$ &  \\
&  & \ \ $\pi /2$ \ \ $\pi /2$ \ \ \ $\pi $ \ \ \ $5\pi /4$ &  \\
&  & \ \ \ \ \ \ \ \ $3\pi /4$ \ \ \ $\pi $ &  \\ \hline
\end{tabular}%
\end{table}

\begin{table}[tbp]
\caption{Schemes of the amplitude and phase profiles corresponding to vortex
solitons with $S=2$.}
\label{tabela2}
\par
\begin{tabular}{|l|l|l|l|}
\hline
Type & $S$ & \ \ Phase configuration & Stable (plane wave/cylindrical wave)
\\ \hline
\ \ \ \ $i$ & \ \ 2 & \ \ \ \ \ \ \ \ $\pi /2$ \ \ \ 0 \ \ $3\pi /2$ & \ \ \
\ No/No \\
&  & \ \ \ \ \ \ \ \ \ $\pi $ \ \ \ \ \ x \ \ \ \ $\pi $ &  \\
&  & \ \ \ \ \ \ \ $3\pi /2$\ \ \ \ 0 \ \ \ $\pi /2$ &  \\ \hline
\ \ \ \ $j$ & \ \ 2 & \ \ \ \ \ \ \ \ \ \ \ \ \ \ \ \ 0 & \ \ \ \ Yes/No \\
&  & \ \ \ \ \ \ \ $\pi /2$\ \ \ \ x\ \ \ \ $3\pi /2$ &  \\
&  & \ $\pi $\ \ \ \ \ \ x\ \ \ \ \ \ x\ \ \ \ \ \ x \ \ \ \ \ \ $\pi $\  &
\\
&  & \ \ \ \ \ \ $3\pi /2$\ \ \ \ x\ \ \ \ \ $\pi /2$ &  \\
&  & \ \ \ \ \ \ \ \ \ \ \ \ \ \ \ \ 0 &  \\ \hline
\ \ \ \ $k$ & \ \ 2 & \ \ \ \ \ \ \ \ \ \ \ \ \ 0 \ \ $\pi $ & \ \ \ \
Yes/Yes \\
&  & \ \ \ \ \ \ \ \ \ \ \ \ \ $\pi $ \ \ 0 &  \\ \hline
\ \ \ \ $l$ & \ \ 2 & \ \ \ \ \ \ \ \ \ \ \ \ 0 \ \ $3\pi /2$ & \ \ \ \ No/No
\\
&  & \ \ \ \ \ $\pi /2$ \ 0 \ \ \ \ $\pi $ \ \ \ \ $\pi $ &  \\
&  & \ \ \ \ \ \ $\pi $ \ \ \ $\pi $ \ \ \ \ 0 \ \ $\pi /2$ &  \\
&  & \ \ \ \ \ \ \ \ \ \ $3\pi /2$ \ 0 &  \\ \hline
\end{tabular}%
\end{table}

Thus, we conclude that the changes in values of phase shifts $\phi _{1,2}$
in Eq. (\ref{eq12}), which are introduced by the oblique incident plane
waves, as well as the detuning, $F$, do not cause qualitative changes in the
structure and effective stability of the 2D self-trapped fundamental modes
and particular types of vortex complexes, of type $a$ with $S=1$, and type $%
k $ with $S=2$ (the quadrupole). Finally, systematic simulations demonstrate
that moving discrete-soliton complexes cannot be found in the present 2D
model.

\subsection{Rabi solitons driven by the cylindrical wave}

For the 2D QD array with the cylindrical-wave excitation, described by Eq. (%
\ref{eq18}), we report here results obtained for $ka=\pi /4$. As we shown by
detailed analysis, these results adequately represent the generic situation.

\begin{figure}[h]
\center\includegraphics [width=12cm]{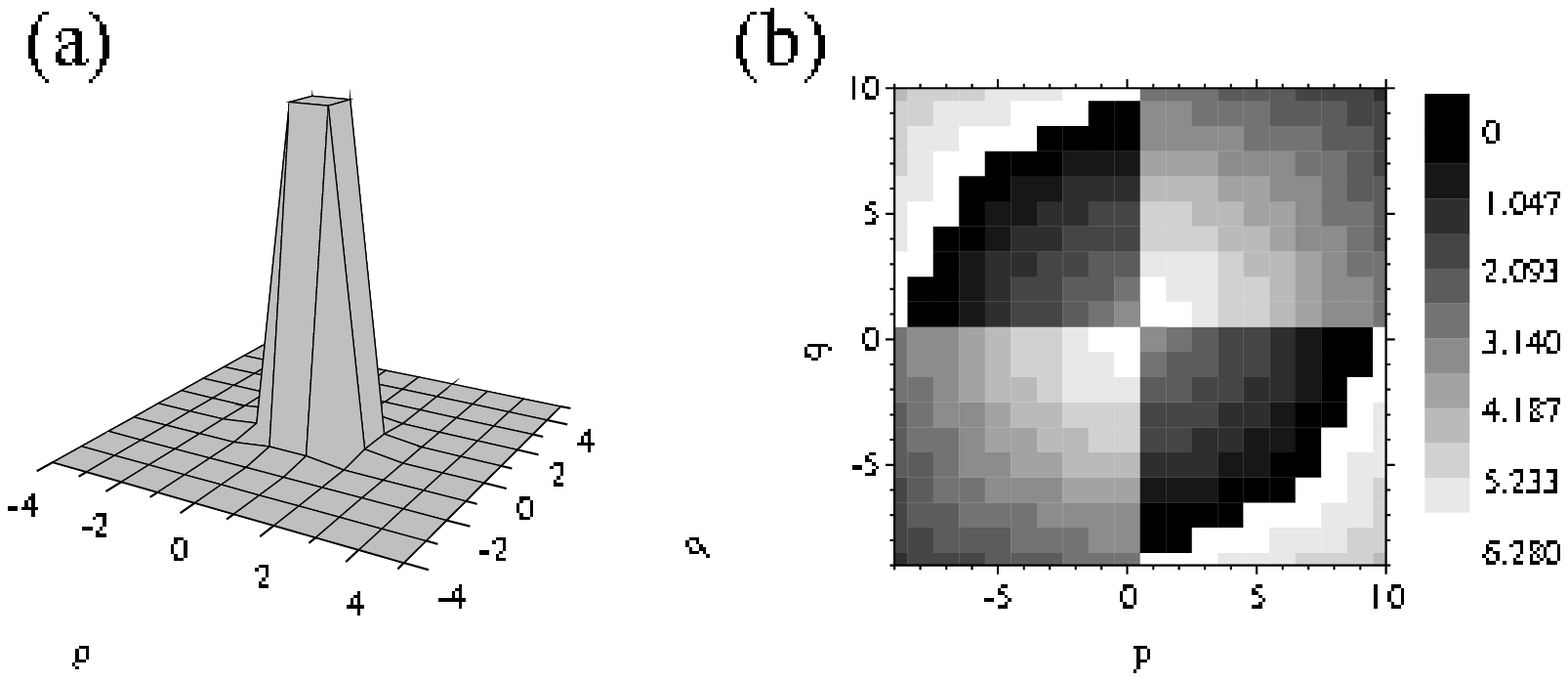}
\caption{The amplitude (a) and phase (b) patterns of the vortex in each
component of the complex of the $k$ type with $S=2$, excited by cylindrical
wave.}
\label{cyl3}
\end{figure}

Similar to the planar-wave model, soliton complexes are located in the
semi-infinite gap in the linear spectrum. However, unlike Eq. (\ref{disp}),
the spectrum with the cylindrical driving wave cannot be found analytically,
therefore it was calculated numerically. Two-component fundamental solitons
are built, as above, of two components, each of them being of the onsite,
inter-site, or hybrid type. Vortex complexes found in the present system are
displayed in Fig. \ref{cyl3}. These are complexes of $b$ and $l$ types with $%
S=1$, or of $k$ and $h$ types with $S=2$, see Tables I and II. In the system
with a nonzero detuning parameter $F\neq 0$, all types of the
above-mentioned fundamental and vortical complexes are generated, but the
symmetry between the two components is not preserved.

\begin{figure}[h]
\center\includegraphics [width=12cm]{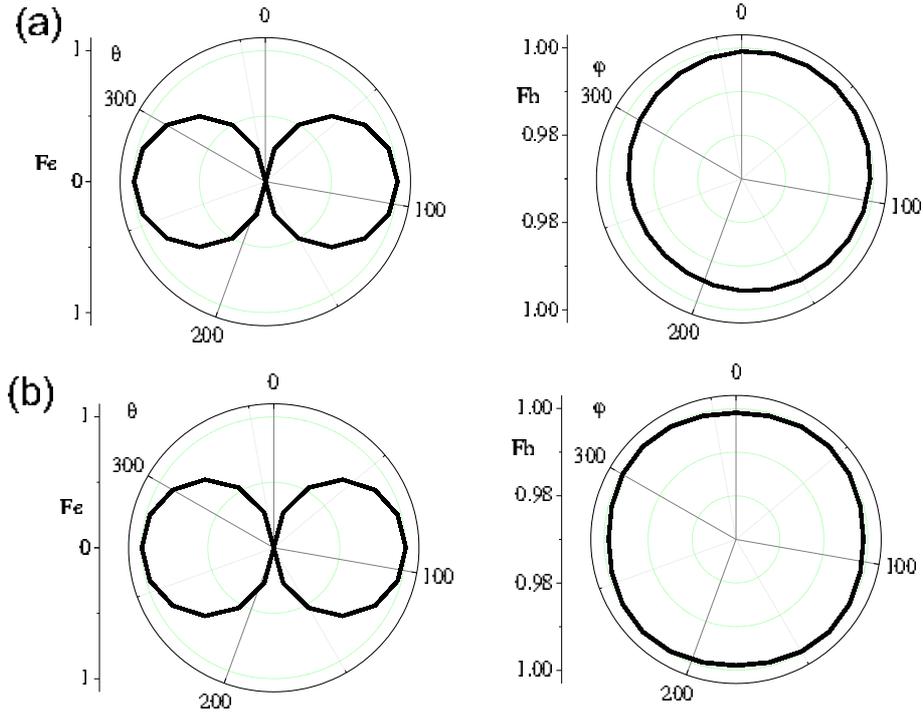}
\caption{Radiation patterns $F_{e}$ and $F_{h}$ for (a) the onsite
complex with $\protect\mu =-17,\,\Delta \protect\omega
=-30,\,F=1$, and (b) the vortical complex of the $k$ type with
$F=0,\protect\mu =17.2,S=2$, in the model with the
cylindrical-wave excitation, Eq. (\protect\ref{eq18}). }
\label{cyl2r}
\end{figure}

The linear-stability analysis shows that the complexes formed
identical fundamental-soliton onsite components are stable for all
value of detuning $F$, and there are stability regions in the
respective parameter space for vortex configuration of type $k$
with $S=2$ at $F=0$. The stability region of latter mode almost
overlaps with its counterpart for the corresponding vortex complex
in the 2D system with the plane-wave excitation. Radiation
patterns of stable solitary Rabi structures excited by the
cylindrical wave are plotted in Fig. \ref{cyl2r}.

\begin{figure}[h]
\center\includegraphics [width=12cm]{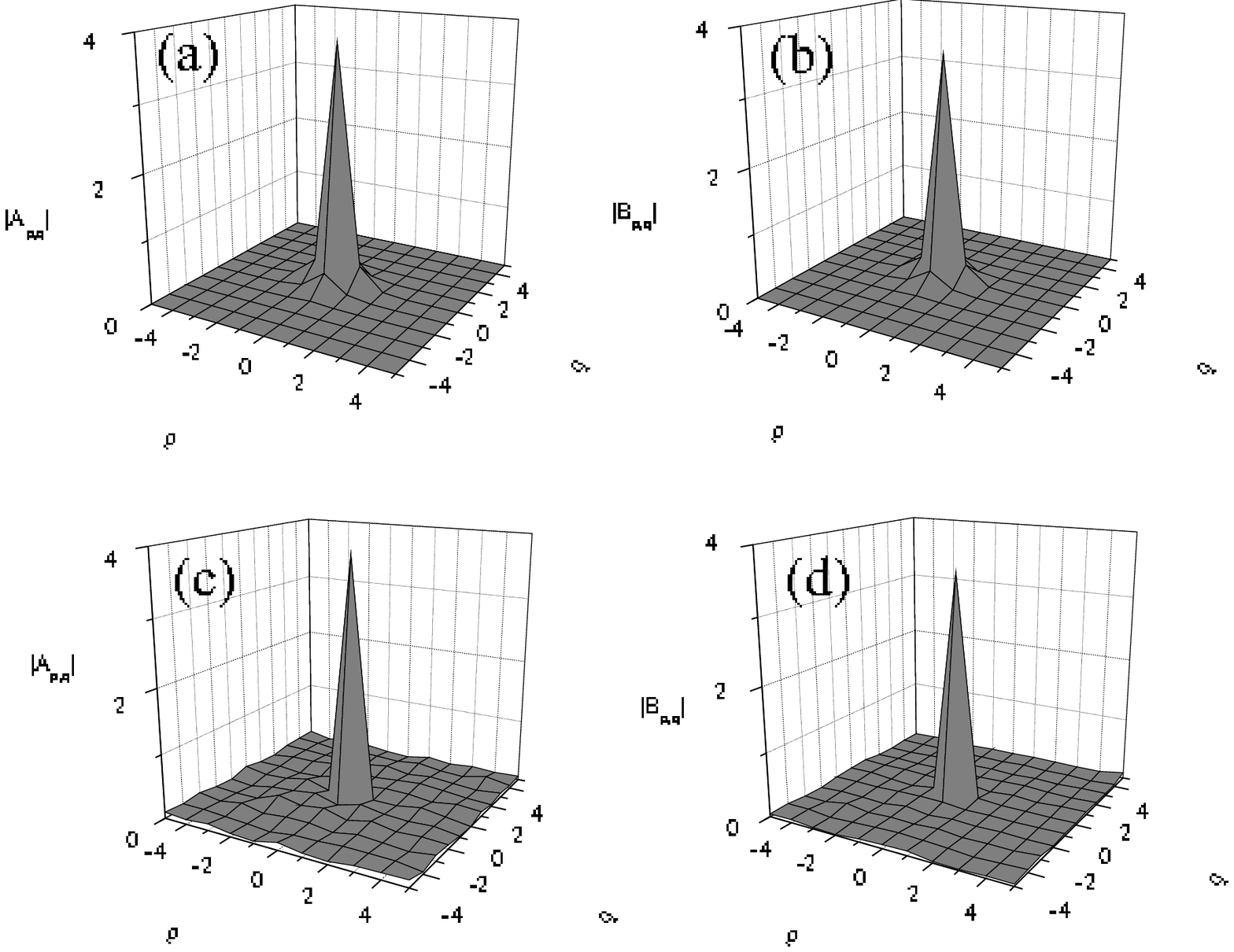}
\caption{The evolution of the amplitude pattern of the onsite soliton
complex with $\protect\mu =-17,\,\Delta \protect\omega =-30,\,F=1$ in the
model with the cylindrical-wave excitation, see Eq. (\protect\ref{eq18}).
Plots correspond to $t=0,\, 50, \, 100$ presented in arbitrary units.}
\label{cyl2}
\end{figure}

\begin{figure}[h]
\center\includegraphics [width=12cm]{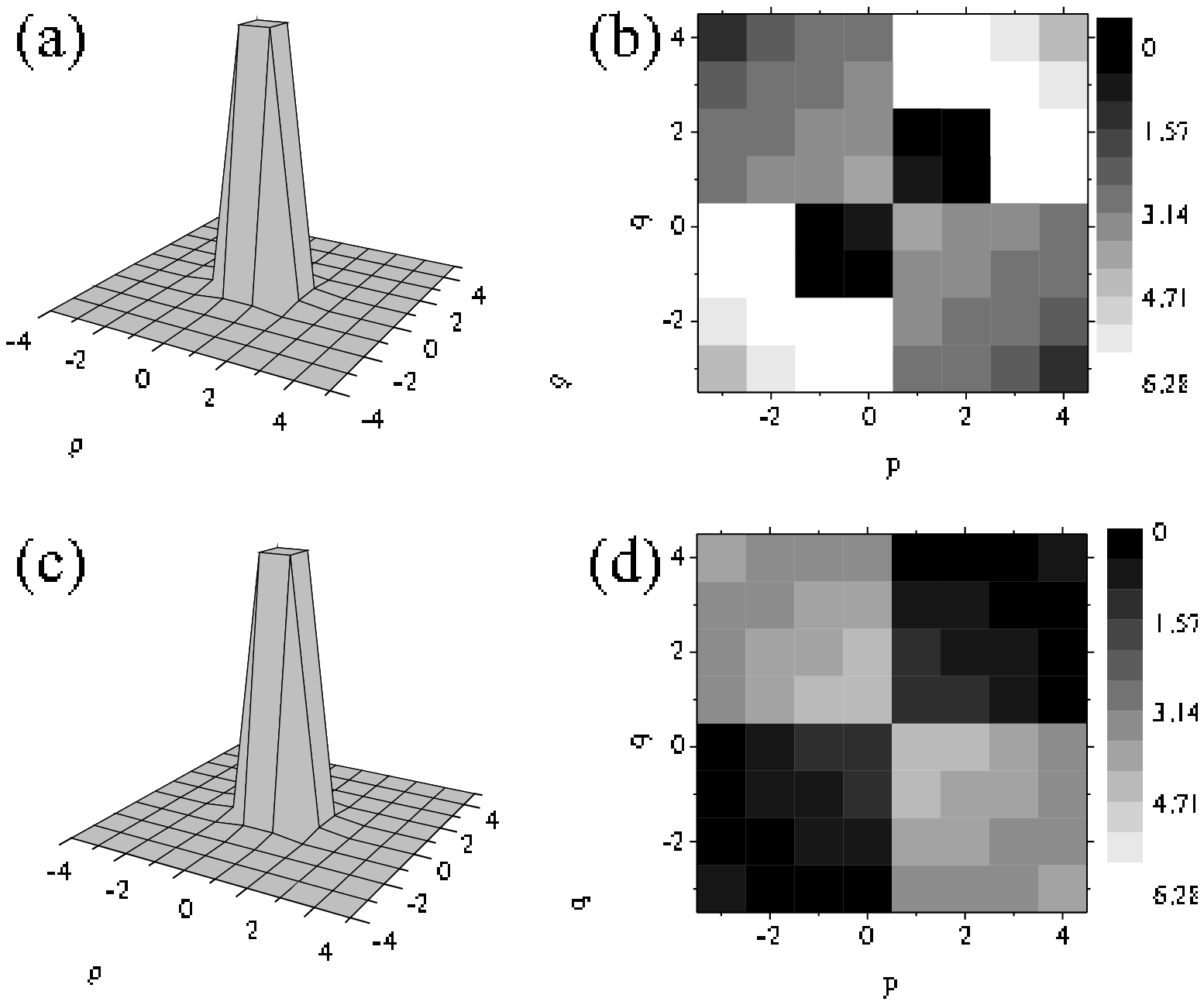}
\caption{The evolution of the stable vortex complex of the $k$ type, excited
by the cylindrical wave, is illustrated by amplitude and phase patterns,
shown at two moments of time. Parameters are $F=0,\protect\mu =17.2,S=2$.
Plots correspond to $t=0$ and $50$ in arbitrary units.}
\label{cyl4}
\end{figure}

\begin{figure}[h]
\center\includegraphics [width=12cm]{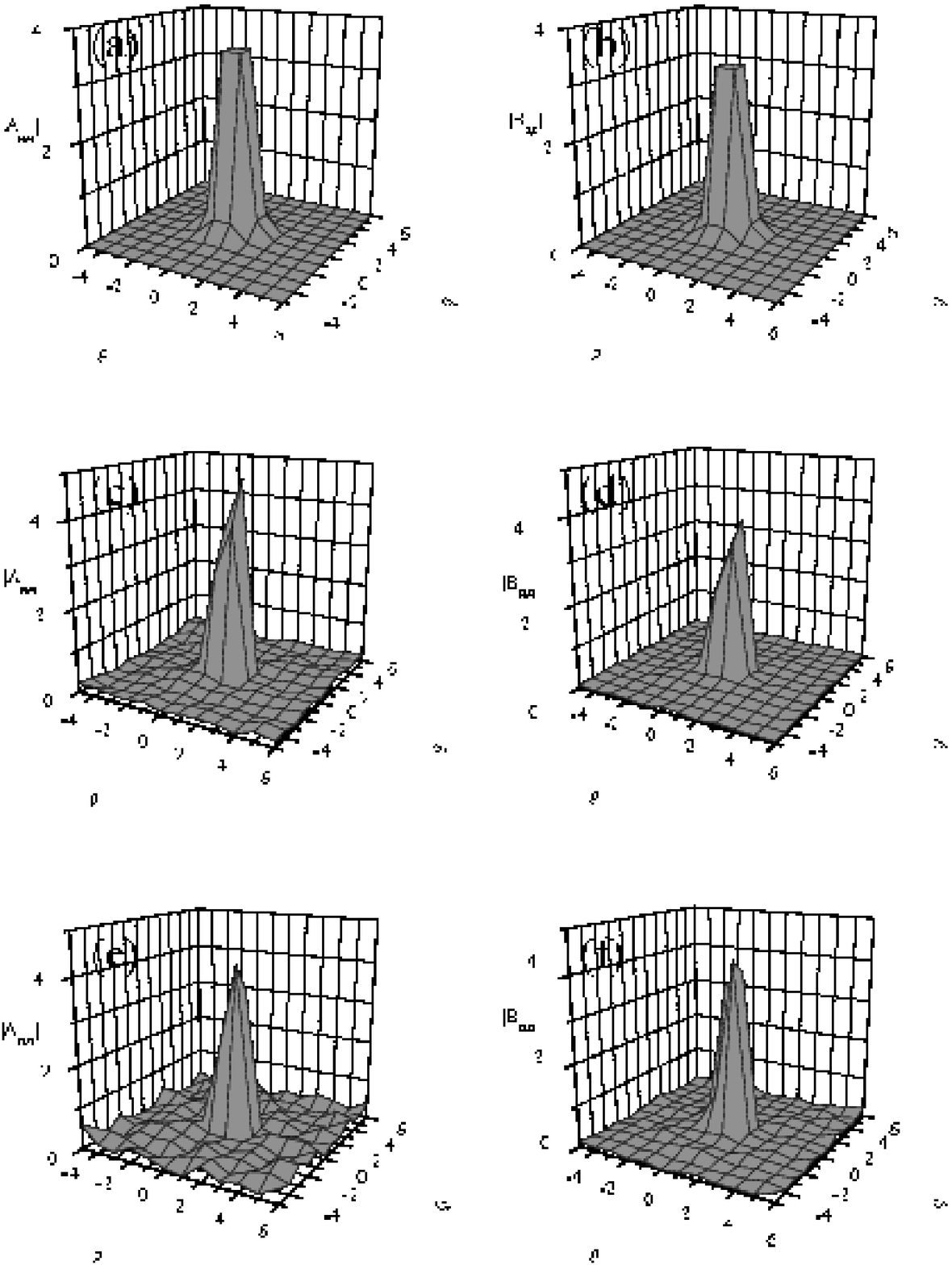}
\caption{The perturbed evolution of the amplitude pattern of an offsite
soliton, for $\protect\mu =-17,\,\Delta \protect\omega =-110,\,F=1$. The
cylindrical-wave excitation is used, see Eqw. (\protect\ref{eq18}). Plots
correspond to $t=0,\, 50, \, 100$ in arbitrary units. }
\label{cyl5}
\end{figure}

\begin{figure}[h]
\center\includegraphics [width=12cm]{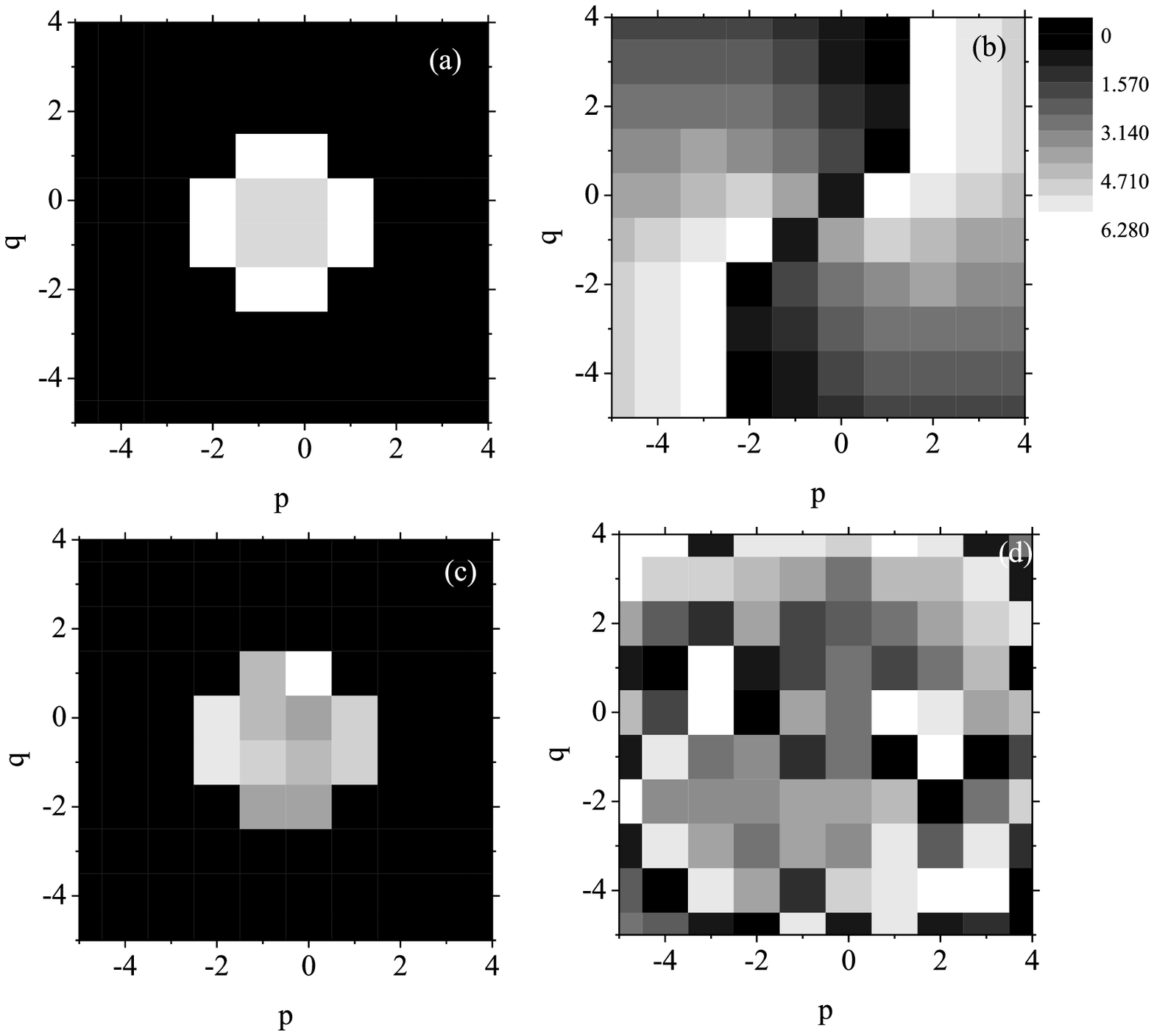}
\caption{The evolution of an unstable vortex of the $l$ type in the model
with the cylindrical-wave excitation. Parameters are $F=0,\protect\mu %
=17.2,S=1$. Plots correspond to $t=0$ and $t=100$ in arbitrary units.}
\label{cyl6}
\end{figure}

The predicted dynamical stability has been confirmed by direct
simulations for the onsite fundamental soliton complexes (Fig.
\ref{cyl2}) and vortex complexes for particular values of the
system parameters. Similar to what was seen above, stable vortices
preserve their amplitude and phase patterns, as shown in Fig.
\ref{cyl4}. In general, effectively stable localized patterns
(including those which are unstable in terms of the eigenvalues
but exhibit persistence in direct simulations) evolve as breathing
modes. In contrast, the unstable hybrid and inter-site complexes
radiate into background a significant part of their energy. The
remaining energy can be reorganized as a new localized onsite
breathing structure, see Fig. \ref{cyl5}. On the other hand,
unstable vortex complexes are more robust structures with
significantly reduced radiation of energy to the background. The
newly formed breathing structures preserve localization but phase
coherence is destroyed. The symmetry of the profile with respect
to the center of vortex is broken, as well as the symmetry between
the components, see Fig. \ref{cyl6}.

In the physical system underlying the present lattice model, parameters may
be known only with a finite accuracy, which makes it necessary to test the
stability of the localized modes with respect to random variations of
coefficients of Eqs. (\ref{eq12}) and (\ref{eq18}). We have performed this
test by introducing random changes, of relative amplitude $10\%$, of
coefficients $\xi $ and $F$, which represent uncertainties in QD energy
levels, and tunneling matrix elements that determine the strength of the
inter-site coupling. The result (not shown here in detail) is that all the
localized modes which were found above to be dynamically stable are also
structurally stable against the random variations of the system's
parameters. In fact, the robustness of discrete solitons against random
deformations of the underlying lattices is known in earlier studied models
\cite{1}.

The bottom line of this subsection is that in the system with the
cylindrical-wave excitation, fundamental onsite soliton complexes and
vortical ones of the $k$ type with $S=2$ are stable in certain areas of
their existence region. In addition, we point out that the soliton-produced
radiation patterns may be used as digital signals for data coding in the
framework of the quantum information transition. Due to the different
patterns of these signals for different types of the models, driven by the
plane and cylindrical waves, they may be identified clearly.

\subsection{The concept of solitonic nano-antennas}

As said above, the transmitting antenna is defined as a device which
converts the near field into the far field \cite{32}. On the other hand, the
Rabi solitons feature the spatially distributed polarization induced by the
RO, as seen in Eqs. (\ref{eq23}) and (\ref{eq25}). The 2D electromagnetic
radiation patterns induced by these polarization profile suggest to consider
the present system in the context of the realization of transmitting
antennas. Thus, the Rabi soliton offers a previously unexplored physical
mechanism to emulate the antennas, while the driving field plays the role of
the external energy source launching and driving this device.

For the data transfer, the emitted field should be temporally modulated by
an input signal through the driving electromagnetic field. Actually, this
modulation is relatively slow, hence it may be considered as an adiabatic
process. As mentioned above, a basic characteristic of the antenna is the 2D
radiation pattern, or, in some cases, partial patterns in the $E$- and $H$%
-planes \cite{32}. The necessity to consider the radiation profiles in 2D
makes the present analysis principally different from the study of RO
solitons in 1D arrays, which was recently reported in Ref. \cite{31}.

Radiation pattern (\ref{antplanar11}) produced by the QD array can be
rewritten in terms of the standard notation adopted in the antenna theory
\cite{32} as $F(\theta ,\varphi )=\sin {(\theta )}\cdot (AF)_{N_{1}\times
N_{2}}$. The first factor, $\sin {(\theta )}$, is the radiation pattern of a
single emitter (in our case, the electric dipole), while the second one is
the array factor,
\begin{eqnarray}
(AF)_{N_{1}\times N_{2}}
=\sum_{p=-N_{1}/2}^{N_{1}/2}\sum_{q=-N_{2}/2}^{N_{2}/2}A_{p,q}B_{p,q}^{\ast
}\exp \left\{ {ip}\left[ {c}^{-1}{\omega a\,}\left( {\sin {\theta }}\right)
\left( {\cos {\varphi }}\right) {+2\phi _{1}}\right] \right\}  \nonumber \\
\times \exp \left\{ {iq}\left[ {c}^{-1}\omega {a\,}\left( {\sin {\theta }}
\right) \left( {\sin {\varphi }}\right) {+2\phi _{2}}\right] \right\} .
\label{kll}
\end{eqnarray}%
In the macroscopic phased antenna array, elements are excited by a signal
having a constant amplitude and the phase which uniformly increases along
the array's axis \cite{32}. The array factor for that case corresponds to
Eq. (\ref{kll}) with $A_{p,q}B_{p,q}^{\ast }=1$. The phase variation allows
one to perform scanning of the space by turning off the main lobe of the
radiation pattern. One of the most efficient means to implement the antenna
control is the use of switched-beam systems which can impose different
angular patterns, in order to enhance the emitted signal in a preferable
direction. The beam-forming algorithm is implemented through a complex
pattern of the excitation of individual elements [which corresponds to
factor $S_{pq}=A_{p,q}B_{p,q}^{\ast }$ in notation (\ref{kll})], adjusted so
as to place the maximum of the main beam in the necessary direction \cite{32}%
.

The purport of the solitonic nano-antenna concept is to introduce the
beam-forming algorithm, which is determined by the set of factors $S_{pq}$,
using the soliton-emission profile. As several different stable solitons may
exist in the given array, the choice of the overall profile depends on
initial conditions. This implies that a finite number of predefined array
factors exist for a given set of antenna parameters, while the initial
conditions give a facility for choosing and switching the suitable one. To
the best of our knowledge, this feature has no analogs in previously
developed types of nano-antennas, and it seems quite promising from the
practical point of view.

For the stable solitons built of two identical components, $A$ and $B$ in
terms of Eq. (\ref{eq21}), the array-factor pattern is independent on the
phase profile of the soliton. Partial radiation patterns in the $E$- and $H$%
-planes for different types of the solitons and different types of the
excitation are displayed in Figs. \ref{vortex1}, \ref{slika1r},\ref{vortex4}%
, and \ref{cyl2r}.

A basic problem that one should take care of to implement emitting
nano-arrays in the real physical setting is maintaining the coherence of
individual emitters in the array over sufficiently long time. The
destruction of the coherence is contributed to by physical mechanisms of
dephasing and relaxation, as well as by material imperfections of the
underlying array. Recent experimental results clearly demonstrate that these
problems may be resolved, allowing quite large QD arrays to maintain the
coherent behavior even at room temperatures \cite{extra1,extra2}.

There are two types of QDs that are commonly employed in nanophotonics \cite%
{extra3}. The first one is built of semiconductor nanocrystals (typically
II-IV compounds) embedded in a glass matrix. In Ref. \cite{extra1}, the QD
ensemble of the CdSe/ZnS core/shell nanocrystals, capped with
octade-cylamine, has been produced. It was fabricated as a 2D
inhomogeneously broadened close-packed network of $50\times 50$ sites, with
the average diameter and interdot distance $5.2$ and $7.9$nm, respectively.
There also appeared dark QDs associated with defects in the ensemble. The
qualitative picture of the exciton dynamics in the QD array observed in Ref.
\cite{extra1}, based on direct measurements, shows that when an exciton is
photoexcited in a high-energy QD, energy transfer occurs preferentially to a
low-energy QD. At low temperatures, the exciton is trapped at a local
low-energy site to which the energy was transferred. In contrast to that, at
room temperature the exciton hops repeatedly, until it is transferred to a
dark QD, where it undergoes nonradiative recombination. When an exciton is
photoexcited in a low-energy QD, it tends to be trapped in it. However, at
room temperature, there is a non-negligible probability for the exciton to
be transferred from a lower-energy QD to a higher-energy one. In Ref. \cite%
{extra1}, time- and spectrally resolved fluorescence intensities were
measured by means of site-selective spectroscopy at both room and low ($80$
K) temperatures. The respective inverse radiative decay rate is found to be $%
15 $ ns, which makes such type of arrays promising for the implementation of
nano-antennas.

The second type of QDs is the self-organized structures produced by the
epitaxial crystal growth in the Stranski-Krastanov regime \cite{extra3}. The
self-assembled QD array was grown in Ref. \cite{extra2} with the help of the
molecular-beam epitaxy on a semiconducting ($100$)-oriented GaAs background,
with a $500$ nm N$^{+}$-GaAs buffer layer. The respective lens-shaped In$%
_{0.65}$Al$_{0.35}$As quantum dot is a part of a sphere with a fixed height
of $3.4$ nm and base diameter of $38$ nm. The photoluminescence spectrum of
the QD array is time-resolved at the excitation density of $1065$ W/cm$^{2}$%
, at $77$ K. The corresponding exciton lifetime, found from the
measurements, is $800-1200$ ps, which complies with the respective value for
the arrays of the first type \cite{extra1}. Note that, for the second type,
the growth of the top layer allows the cavities to be formed and the
electrical contacts to be applied, which makes it especially suitable for
the implementation of nano-antennas.

Lastly, it is relevant to mention that the soliton mechanism of the
self-trapping of the antenna helps to mitigate detrimental effects of
imperfections of the underlying QD lattice, as discrete solitons are well
known to be robust against deformations of the lattice \cite{1}. Moreover,
the fact that the phase structure of topologically organized solitons, such
as discrete vortices \cite{vortex}, is stable in imperfect lattices, also
indicates that the so constructed nano-antennas tend to stabilize themselves
against dephasing.

It is relevant to note that the self-organized lattice built of
semiconductor QDs not a unique structure allowing the
implementation of soliton-based nano-antennas, and the RO is not a
unique enabling mechanism for this. Another promising way for
achieving this purpose is suggested by theoretical analysis of
discrete dissipative plasmon solitons in an array of graphene QDs
\cite{2new}.The single QD in such an array represents a doped
graphene nanodisk placed on top of the plane background. The
single-dot excitation represents a confined surface plasmon with
the resonant frequency in the THz or infrared range. The QDs in
the array are coupled by long-range dipole-dipole interactions. As
it was demonstrated in \cite{2new}, the soliton formation and its
stability take place under the control of an incident driving
electromagnetic wave via the Kerr optical nonlinearity. In spite
of the different physical origins, both the surface plasmon
solitons predicted in \cite{2new} and Rabi solitons considered
above exhibit strongly confined one- or two-peak areas of
electrical polarization. Such peaks may be symmetric or
asymmetric, while the spatial structure of the confinement area is
tunable via the incidence angle of the oblique driving field.
Thus, the above-mentioned graphene QD-array makes it possible to
design an electrically controlled nano-antenna for THz and
infrared frequency ranges.
The radiation pattern (or array factor) for this antenna is given by Eq. (%
\ref{kll}) with necessary modifications (the slowly varying
amplitudes of orthogonally directed dipole moments can be found as
a solutions of the nonlinear coupled equations derived in
\cite{2new}). The qualitative shape of such a radiation pattern is
similar to that of the Rabi-soliton-based nano-antenna proposed
above.

\section{Conclusions}

The objective of this work is to introduce the concept of the tunable
nano-antenna array based on the discrete-soliton patterns formed in the 2D
nonlinear lattices of semiconductor QDs (quantum dots). These lattices can
be realized as square-shaped arrays of identical two-level quantum
oscillators (the self-organized semiconductor QDs), coupled to nearest
neighbors by the electron-hope tunneling and interacting with the external
electromagnetic field. The local-field corrections, which account for the
difference between the field inside the QD and the external field, induce
the nonlinearity of the electron-hole motion inside each QD.

The main conclusions of our study are summarized as follows.

(i) The model of RO (Rabi oscillations) in the 2D QD lattice has been
derived, taking the inter-dot tunnelling and local-field correction into
account. The model is based on a set of linearly and nonlinearity coupled
DNLS equations for probability amplitudes of the ground and first excited
states of two-level oscillators (QDs). Two different driving electromagnetic
fields were considered, \textit{viz}., the plane-wave and cylindrical ones.
In the former case, the coupling coefficients are complex, with absolute
values independent of the QD position, and phases linearly increasing across
the QD array in both directions. For the cylindrical-wave drive, the
absolute values of the coupling coefficients depend on the distance between
the given QD and the source of the driving field. The corresponding phases
depend nonlinearly on the distance.

(ii) Stable discrete-soliton complexes are found. They are built of onsite
fundamental single-peaked solitons, taken in both components, or discrete
vortex solitons of certain types.

(iii) The emission properties of stable solitary modes have been
characterized by angular radiation patterns. These patterns strongly depend
on the type of the discrete localized mode.

(iv) The concept of self-assembling nano-antennas, based on the stable
discrete-soliton complexes in the nonlinear lattices, is introduced. The
necessary type of the localized mode may be selected by the initial
conditions, which, in turn, can be controlled by the external optical field
(to be supplied in the form of a strong laser pulse). As a consequence, a
finite number of predefined radiation patterns can be provided by a given
antenna. This way of the operational control of nano-antennas has no analogs
in previously developed antenna schemes. The stability of the self-trapped
nano-antennas against structural imperfections and intrinsic dephasing has
been considered. Thus, the system proposed here can be related to
switched-beam systems used in macroscopic antennas \cite{32}, which is a
promising setting for applications to nanoelectronics and nanooptics.

(v) The concept of soliton-excited nano-antennas may be carried
over to the surface-plasmon mechanism of the soliton formation in
the array of graphene QDs with the Kerr nonlinearity predicted in
\cite{2new}. This type of the nano-antennas is promising for
applications in the THz and infrared frequency ranges.

\section*{ Acknowledgements}

G.G., A.M., and Lj.H. acknowledge support from the Ministry of Education and
Science of Serbia (Project III45010). G.S. acknowledges support from the EU
FP7 projects FP7 People 2009 IRSES 247007 CACOMEL and FP7 People 2013 IRSES
612285 CANTOR.


\section*{References}

\begin{thebibliography}{99}

\bibitem{optics-review} Lederer F, Stegeman G I, Christodoulides D N,
Assanto G, Segev M and Silberberg Y 2008 \textit{Phys. Rep.} \textbf{463} 1

\bibitem{BEC-original} Trombettoni A and Smerzi A 2001 \textit{Phys. Rev.
Lett.} \textbf{86} 2353; Abdullaev F Kh, Baizakov B B, Darmanyan S A,
Konotop V V and Salerno M 2001 \textit{Phys. Rev. A} \textbf{64} 043606;
Alfimov G L, Kevrekidis P G, Konotop V V, and Salerno M 2002 \textit{Phys.
Rev. E} \textbf{66} 046608; Carretero-Gonz\'{a}lez R and Promislow K 2002
\textit{Phys. Rev. A} \textbf{66} 033610; Efremidis N K and Christodoulides
D N 2003 \textit{ibid}. \textbf{67} 063608; Jin G-R, Kim C-K, Nahm K 2005
\textit{ibid}. \textbf{72} 045601; Maluckov A, Had\v{z}ievski Lj, Malomed B
A and Salasnich L 2008 \textit{ibid}. \textbf{78} 013616

\bibitem{BH} Fisher M P A, Grinstein G and Fisher D S 1989 \textit{Phys.
Rev. B} \textbf{40} 546; Greiner M, Mandel O, Esslinger T, Hansch T and
Bloch I 2001 \textit{Nature} \textbf{415} 39

\bibitem{BH-review} Giamarchi T 2004 \textit{Quantum Physics in One Dimension%
} (Oxford University Press, Oxford); Bloch I, Dalibard J and Zwerger W 2008 {%
\textit{R}ev. Mod. Phys.} \textbf{80} 885; Cazalilla M A, Citro R, Giamarchi
T, Orignac E and Rigol M 2011 \textit{\ ibid}. \textbf{83} 1405

\bibitem{Maciek} Lewenstein M, Sanpera A and Ahufinger V 2012 \textit{%
Ultracold Atoms in Optical Lattices: Simulating Quantum Many-Body Systems}
(Oxford University Press, Oxford)

\bibitem{1} Kevrekidis P G 2009 \textit{The Discrete Nonlinear Schr\"{o}%
dinger Equation: Mathematical Analysis, Numerical Computations, and Physical
Perspectives} (Springer, Berlin)

\bibitem{5} Deconinck B, Kevrekidis P G, Nistazakis H E and Frantzeskakis D
J 2004 \textit{Phys. Rev. A} \textbf{70} 063605

\bibitem{6} Ballagh R J, Burnett K and Scott T F 1997 \textit{Phys. Rev.
Lett.} \textbf{78} 1607

\bibitem{7} Ohberg P and Stenholm S 1999 \textit{Phys. Rev. A} \textbf{59}
3890

\bibitem{8} Williams J, Walser R, Cooper J, Cornell E and Holland M 1999 {%
\textit{P}hys. Rev. A} \textbf{59} R31

\bibitem{Herring} Herring G, Kevrekidis P G, Malomed B A, Carretero-Gonz\'{a}%
lez R and Frantzeskakis D J 2007 \textit{Phys. Rev. E} \textbf{76} 066606

\bibitem{16} Engheta N, Salandrino A and Al\`{u} A 2005 \textit{Phys. Rev.
Lett.} \textbf{95} 095504

\bibitem{17} \textit{Quantum Dot Lasers} 1998 (edited by M. G. D. Bimberg
and N. N. Ledentsov, Wiley, Chichester, England)

\bibitem{18} Novotny L and Hecht B 2006 \textit{Principles of Nano-Optics}
(Cambridge University Press, New-York)

\bibitem{19} Scully M O and Zubairy M S 2001 \textit{Quantum Optics}
(Cambridge University Press, Cambridge, England)

\bibitem{20} Xie Q-T, Cui S, Cao J-P, Amico L and Fan H 2014 \textit{Phys.
Rev. X} \textbf{4} 021046

\bibitem{21} Slepyan G Ya, Maksimenko S A, Hoffmann A and Bimberg D 2002
\textit{Phys. Rev. A} \textbf{66} 063804

\bibitem{22} Slepyan G Ya and Maksimenko S A 2008 {\textit{N}ew J. Phys.}
\textbf{10} 023032

\bibitem{23} Slepyan G Ya, Magyarov A, Maksimenko S A, Hoffmann A and
Bimberg D 2004 {\textit{P}hys. Rev. B} \textbf{70} 045320

\bibitem{24} Slepyan G Ya, Magyarov A, Maksimenko S A and Hoffmann A 2007 {%
\textit{P}hys. Rev B.} \textbf{76} 195328

\bibitem{25} Kibis O V, Slepyan G Ya, Maksimenko S A and Hoffmann A 2009 {%
\textit{P}hys. Rev. Lett.} \textbf{102} 023601

\bibitem{26} Slepyan G Y, Yerchak Y, Maksimenko S A and Hoffmann A 2009 {%
\textit{P}hys. Lett. A} \textbf{373} 1374

\bibitem{27} Slepyan G Y, Yerchak Y D, Hoffmann A and Bass F G 2010 \textit{%
Phys. Rev. B} \textbf{81} 085115

\bibitem{28} Slepyan G Y, Yerchak Y D, Maksimenko S A, Hoffmann A and Bass F
G 2012 \textit{Phys. Rev. B} \textbf{85} 245134

\bibitem{29} Mitsumori Y, Hasegawa A, Sasaki M, Maruki H and Minami F 2005
\textit{Phys. Rev. B} \textbf{71} 233305

\bibitem{30} Asakura K, Mitsumori Y, Kosaka H, Edamatsu K, Akahane K,
Yamamoto N, Sasaki M and Ohtani N 2013 \textit{Phys. Rev. B} \textbf{87}
241301(R)

\bibitem{31} Gligori\'{c} G, Maluckov A, Had\v{z}ievski Lj, Slepyan G Ya and
Malomed B A 2013 \textit{Phys. Rev. B} \textbf{88} 155329



\bibitem{32} Balanis C A 1997 \textit{Antenna Theory: Analysis and Design}
(New York: Wiley)

\bibitem{33} Biagioni P, Huang Y-S and Hecht B 2012 \textit{Rep. Prog. Phys.}
\textbf{75} 024402

\bibitem{34} Novotny L and van Hulst N 2011 \textit{Nature Photonics}
\textbf{5} 83

\bibitem{35} Engheta N 2007 \textit{Science} \textbf{317} 1698

\bibitem{36} Hanson G W 2005 \textit{IEEE Trans. Antennas Propag.} \textbf{53%
} 3426

\bibitem{37} Burke P J, Li S and Yu Z 2006 \textit{IEEE Trans. Nanotechnol.}
\textbf{5} 314

\bibitem{38} Slepyan G Ya, Shuba M V, Maksimenko S A and Lakhtakia A 2006
\textit{Phys. Rev. B} \textbf{73} 195416

\bibitem{39} Ren L, Zhang Q, Pint C L, W\'{o}jcik A K, Bunney M, Arikawa Jr
T, Kawayama I, Tonouchi M, Hauge R H, Belyanin A A and Kono J 2013 \textit{%
Phys. Rev. B} \textbf{87} 161401(R)

\bibitem{40} Novotny L 2007 \textit{Phys. Rev. Lett.} \textbf{98} 266802

\bibitem{41} Al\`{u} A and Engheta N 2008 \textit{Phys. Rev. Lett.} \textbf{%
101} 043901

\bibitem{42} Jornet J M and Akyildiz I F 2013 \textit{IEEE J. Selected Areas
in Communications/ Supplement p.2} \textbf{31} 685

\bibitem{43} Slepyan Gr Y and Boag A. 2013 \textit{Phys. Rev. Lett.} \textbf{%
111} 023602

\bibitem{51} Landau L D and Lifshitz E M 1980 \textit{Statistical Physics}
(Pergamon, New York).


\bibitem{44} Mokhlespour S, Haverkort J E M, Slepyan G, Maksimenko S and
Hoffmann A 2012 \textit{Phys. Rev. B} \textbf{86} 245322



\bibitem{3new} Zheng Z, Zhao C, Lu S, Chen Yu, Li Y, Zhang H
and Wen S 2012 \textit{Optics Express} \textbf{20} 23201

\bibitem{4new} Zhang H, Virally S, Bao Q, Ping L K, Massar S, Godbout N and
Kockaert P 2012 \textit{Opt. Lett.} \textbf{37} 1856

\bibitem{1new} Asakura K, Mitsumori Y, Kosaka H, Edamatsu K, Akahane K,
Yamamoto N, Sasaki M, and Ohtani N 2013 \textit{Phys. Rev}. B
\textbf{87}, 241301(R)



\bibitem{45} Citrin D S 1993 \textit{Phys. Rev. B} \textbf{48} 2535

\bibitem{46} Slepyan G Ya, Maksimenko S A, Lakhtakia A, Yevtushenko O and
Gusakov A V 1999 \textit{Phys. Rev. B} \textbf{60} 17136

\bibitem{47} Gligori\'{c} G, Maluckov A, Step\'{\i}c M, Hadzievski Lj and
Malomed B A 2010 \textit{Phys. Rev. A} \textbf{81} 013633

\bibitem{48} Gligori\'{c} G, Maluckov A, Step\'{\i}c M, Hadzievski Lj and
Malomed B A 2010 \textit{J. Phys. B: At. Mol. Opt. Phys.} \textbf{43} 055303

\bibitem{49} \"{O}ster M and Johansson M 2006 \textit{Phys. Rev. E} \textbf{%
73} 066608

\bibitem{50} Kevrekidis P G, Rasmussen K O and Bishop A R 2000 \textit{Phys.
Rev. E} \textbf{61} 2006

\bibitem{vortex} Malomed B A and Kevrekidis P G 2001 \textit{Phys. Rev. E}
\textbf{64}  026601




\bibitem{extra1} Miyazaki J and Kinoshita S 2012 \textit{Phys. Rev. B}
\textbf{86} 035303

\bibitem{extra2} Ding C R, Wang H Z and Xu B 2005 \textit{Phys. Rev. B}
\textbf{71} 085304

\bibitem{extra3} Fox M 2012 \textit{Quantum Optics: An Introduction} (Oxford
Master Series: Oxford University Press)

\bibitem{2new} Smirnova D A, Noskov R E, Smirnov L A and Kivshar Y S 2015
\textit{Phys. Rev}. B \textbf{91} 075409

\end{thebibliography}
\end{document}